\documentclass[pra,aps,longbibliography,showpacs,groupedaddress,superscriptaddress,twocolumn,toc=flat,nofootinbib]{revtex4-1}
\usepackage{graphicx}
\usepackage{latexsym}
\usepackage{amsmath}
\usepackage{amssymb}
\usepackage{amsfonts}
\usepackage{bm}
\usepackage{bbm}
\usepackage{newtxtext}
\usepackage[varvw]{newtxmath}
\usepackage{subfigure}
\usepackage{verbatim}
\usepackage{siunitx}
\usepackage[unicode=true,
 bookmarks=false,
 breaklinks=false,pdfborder={0 0 1},backref=false,colorlinks=true]
 {hyperref}
\hypersetup{
 linkcolor=[rgb]{0,0,1},citecolor=[rgb]{0,0,1},urlcolor=[rgb]{0,0,1}}
\usepackage{multirow}
\usepackage{color}
\usepackage{comment}
\usepackage{xcolor}
\usepackage{tikz}
\usepackage{tabularx}
\usepackage{enumitem}   
\usepackage{needspace}

\allowdisplaybreaks

\newcommand{\mysection}[1]{%
  \Needspace{6\baselineskip}
  \vspace{1.0ex}
  \begin{center}
    \normalsize\bfseries #1
  \end{center}
  \vspace{-0.3ex}
}

\newcommand{\Ztwo}{$\mathbb{Z}_2$}

\usepackage{braket}

\begin{document}

\newcommand{\mytitle}{Probing quantum spin liquids with multiple quantum coherences}
\title{\mytitle}

\author{Lukas~Homeier}
\email{lukas.homeier@jila.colorado.edu}
\affiliation{JILA and National Institute of Standards and Technology, University of Colorado, Boulder, CO, 80309, USA}
\affiliation{Department of Physics and Center for Theory of Quantum Matter, University of Colorado, Boulder, CO, 80309, USA}

\author{Simon~M.~Linsel}
\affiliation{Department of Physics and Arnold Sommerfeld Center for Theoretical Physics (ASC), Ludwig Maximilian University of Munich, 80333 Munich, Germany}
\affiliation{Munich Center for Quantum Science and Technology (MCQST), 80799 Munich, Germany}


\author{Lode~Pollet}
\affiliation{Department of Physics and Arnold Sommerfeld Center for Theoretical Physics (ASC), Ludwig Maximilian University of Munich, 80333 Munich, Germany}
\affiliation{Munich Center for Quantum Science and Technology (MCQST), 80799 Munich, Germany}

\author{Ana~Maria~Rey}
\email{arey@jila.colorado.edu}
\affiliation{JILA and National Institute of Standards and Technology, University of Colorado, Boulder, CO, 80309, USA}
\affiliation{Department of Physics and Center for Theory of Quantum Matter, University of Colorado, Boulder, CO, 80309, USA}

\date{\today}
\begin{abstract}
Quantum simulators are beginning to prepare long-sought phases of matter that remain difficult to realize cleanly in materials. Yet identifying such phases remains a major challenge when their defining properties are  inherently non-local and cannot be captured by conventional local measurements. Here we establish multiple-quantum coherences (MQCs) as phase-sensitive diagnostics for gapped \Ztwo{}~quantum spin liquids. Focusing on the extended toric code and using large-scale quantum Monte Carlo simulations, we build a direct correspondence between the elementary anyonic excitations and the weights of different MQC sectors. MQCs thereby reveal anyon condensation across the phase transitions through characteristic signatures that remain robust against fluctuations that obscure conventional diagnostics.
Subsystem-resolved MQCs further provide a bipartite entanglement witness that distinguishes a classical loop gas from a quantum-coherent closed-loops gas.
We develop and benchmark a practical protocol to extract MQCs based on the return fidelity of adiabatic round trips, and show that local coherence measurements retain experimentally accessible signatures of the phase transitions. Our results establish MQCs as a practical diagnostic for the excitations, phase structure and global constraints of quantum spin liquids.
\end{abstract}
\maketitle

\mysection{I. Introduction}

Most conventional phases of matter can be identified by measuring an order parameter that reveals a broken symmetry. Quantum spin liquids (QSLs) are a notable exception to this paradigm. Their spins remain disordered even at very low temperatures, while quantum correlations organize the system into a long-range entangled quantum state~\cite{Savary2016,Broholm2020}. In gapped QSLs, this organization supports anyonic excitations and encodes information globally, establishing close connections to topological quantum computing and quantum error correction~\cite{Wen2007,Chen2010,Kitaev2003,Nayak2008}. These properties make QSLs highly appealing, but also exceptionally difficult to identify: no simple local measurement can reveal the character of the phase.

In solid-state materials, evidence for QSL behaviour is typically inferred by combining measurements of magnetic correlations, excitation spectra and transport. Although these measurements have revealed compelling signatures in several candidate materials, their interpretation can depend on detailed modelling. Analog quantum simulators provide a new opportunity: programmable interactions enable the preparation of candidate QSLs, while single-site readout offers access to observables unavailable in most materials. Recent experiments have demonstrated this potential across a growing range of spin-liquid states~\cite{Semeghini2021,Satzinger2021,Bornet2026,Geim2026,Karch2026}.

Greater control and measurement access, however, do not by themselves solve the challenge of identification. The defining structure of a QSL is non-local and must therefore be inferred from correlations involving many particles. Existing non-local observables can provide important signatures, but may be obscured by fluctuations and experimental noise or offer only a partial view of the phase diagram~\cite{Gregor2011,Semeghini2021,Verresen2021,Wang2025}. A broadly useful diagnostic should distinguish the QSL from nearby phases within a common framework while also revealing the excitations responsible for their different character.

\begin{figure*}[t!!]
\centering
\includegraphics[width=\linewidth]{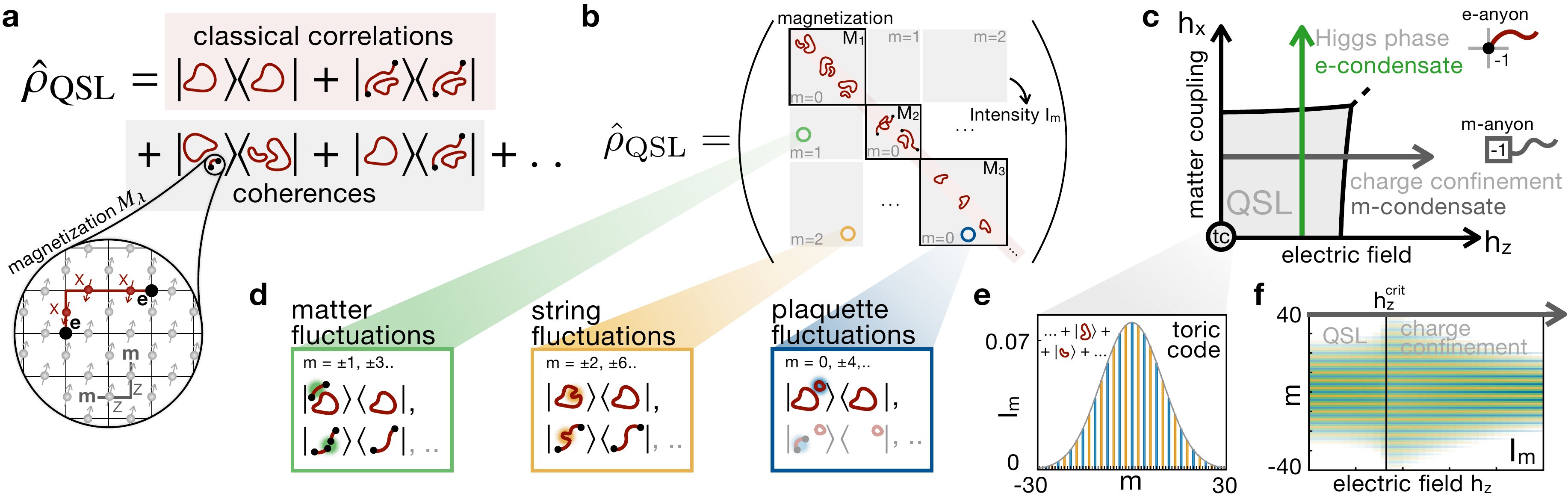}
\caption{\textbf{Multiple quantum coherences in quantum spin liquids.}
\textbf{a} The density matrix~$\hat{\rho}_{\rm QSL}$ describes a system of spins comprised of classical correlations (Gauss' law) and many-body coherences. Inset: Red lines illustrate electric field lines ($\hat{\sigma}^z=-1$). For open strings, the endpoints (black dots) are interpreted as matter excitations or electric e-anyons, created via a string of $\hat{\sigma}^x$ operators. In the conjugate basis, applying a string of $\hat{\sigma}^z$ operators creates magnetic m-anyons.
\textbf{b} The density matrix can be reorganized by sectors of total magnetization~$M_\lambda$, i.e., the total content of electric field lines. The off-diagonal blocks~$m=M_{\lambda}-M_{\lambda'}$ contain coherences between two magnetization sectors -- the multiple quantum coherences (MQCs) with intensity~$I_m$.
\textbf{c} We investigate the MQCs across the extended toric-code phase diagram. By coupling the toric code QSL to external fields~$h_z$ ($h_x$), m-anyons (e-anyons) begin to fluctuate and eventually condense into a trivial phase.
\textbf{d} The sectors~$m$ connect states differing by $m$~spin flips. We can associate the dominant fluctuations in different $m$~sectors with open string (matter) or loop fluctuations.
\textbf{e} The toric-code wavefunction is an equal-weight superposition of closed-loop configurations, yielding a coherence spectrum with non-zero weight only for even~$m$ (orange and blue sectors).
\textbf{f} The coherences~$I_m$ are directly sensitive to m-anyon fluctuations, resulting in a pronounced redistribution of the coherence spectrum across the condensation transition.
}
\label{figure-1}
\end{figure*}

Coherence offers a natural route to such a diagnostic. Atomic clocks and many quantum sensors, for example, use coherence as a precise probe by converting an accumulated phase in coherent single-particle superpositions into a directly measurable population. Extending this idea to many-body states requires characterizing not one, but many different coherences. Recent experiments have begun to access such many-body coherence through round-trip interferometric protocols in dynamically prepared U(1) QSLs~\cite{Karch2026}, motivating a more systematic framework for interpreting such coherence measurements.

Multiple quantum coherences (MQCs), developed in nuclear magnetic resonance~\cite{Baum1985}, provide a simple way to organize this information accessible through phase-sensitive measurements. After choosing a spin-projection axis, MQCs group the off-diagonal elements of the density matrix according to the difference~$m$ in total magnetization between the configurations they connect. The single-quantum sector~$|m|=1$ corresponds to a coherence between two states connected by a single spin flip, while higher orders connect sectors separated by progressively larger changes~$|m|> 1$, extending to the maximal coherence of a Schr{\"o}dinger-cat state. The resulting MQC spectrum thus provides a compact description of many-body coherence without requiring reconstruction of the full quantum state.

The toric code provides a concrete setting in which to explore whether the MQC spectrum can diagnose a QSL. Here, spins reside on the links of a lattice, and sequences of flipped spins (down) form strings that represent emergent electric-field lines, illustrated in red in Fig.~\ref{figure-1}a. In the toric-code ground state, these strings form closed loops, whereas the endpoints of fluctuating open strings mark an electric-anyon (e-anyon) excitation. The ground state is a coherent superposition of all possible closed-loop configurations, and its density matrix contains coherences between the different loop patterns. Here we take advantage of this structure and the way MQCs organize its coherences, and show that the MQC spectrum captures both the e-anyonic excitations and the defining global closed-loop constraint of the QSL, allowing us to distinguish the spin liquid from neighbouring phases. We thereby establish MQCs as phase-sensitive diagnostics of gapped QSLs.

To map these signatures across a full phase diagram, we study the extended toric code, in which the QSL is connected to neighbouring topologically-trivial phases through the condensation of different types of anyons. Within this setting, we identify a direct correspondence between the anyon excitations and the distribution of coherence across the MQC spectrum. As different types of anyons condense, we find characteristic changes in the MQC spectrum that provide robust signatures of the phase boundaries. Using large-scale quantum Monte Carlo simulations, we track these signatures with increasing system size and demonstrate that they develop universal features at the phase transitions. We also construct a witness for bipartite entanglement between subsystems, which share a global closed loop constraint.
Finally, we develop and benchmark a practical protocol for extracting MQCs from the return fidelity of adiabatic round trips~\cite{LewisSwan2020}, avoiding the need to implement time-reversed dynamics. We further demonstrate that MQCs restricted to local subsystems retain clear signatures of the phase transitions while offering greater resilience to decoherence. Together, these results establish MQCs as experimentally accessible probes of the excitations, phase structure and non-local constraints of quantum spin liquids.

\mysection{II. Multiple quantum coherences}

We consider $N$ spin-$1/2$ degrees of freedom, which in the toric-code setting we study, reside on the links of a two-dimensional lattice, see Fig.~\ref{figure-1}a. The definition of MQCs, however, applies to any spin system once a reference spin-projection axis has been chosen. Taking this axis to be $z$, the local Pauli operator $\hat{\sigma}^z_j$ has eigenstates $\ket{\uparrow}_j$ and $\ket{\downarrow}_j$ with eigenvalues $+1$ and $-1$, respectively. A many-body basis state can then be written as a spin configuration $\ket{\lambda}=\ket{\sigma_1,\ldots,\sigma_N}$ and has a definite total spin projection under $\hat{S}^z=\frac{1}{2}\sum_j\hat{\sigma}^z_j$,
such that $\hat{S}^z\ket{\lambda}=M_\lambda\ket{\lambda}$. With this convention, flipping one spin changes the total spin projection by one.

The MQC spectrum organizes the coherences of the many-body state with respect to this collective spin projection or magnetization. To see this, we decompose the density matrix as
\begin{equation}
    \hat{\rho}=\sum_m\hat{\rho}_m,\quad 
    \hat{\rho}_m
    =\sum_{M_\lambda-M_{\lambda'}=m}
    \rho_{\lambda\lambda'}\ket{\lambda}\!\bra{\lambda'}.
\end{equation}
Each component $\hat{\rho}_m$ collects all density-matrix elements connecting configurations whose total magnetization differs by~$m$, as illustrated in Fig.~\ref{figure-1}b. The index~$m$ is referred to as the coherence order. In particular, $m=0$ contains populations and coherences between states with the same total magnetization, whereas $m\neq 0$ resolves coherences between different magnetization sectors.

The different coherence orders can be spectroscopically resolved through a collective phase rotation,
\begin{equation} \label{rot}
    \hat{W}(\phi)=e^{i\phi\hat{S}^z}.
\end{equation}
Because every matrix element within $\hat{\rho}_m$ connects states whose spin projections differ by the same amount, the entire component acquires the same phase $\hat{\rho}(\phi) =\hat{W}(\phi)\hat{\rho}\hat{W}^{\dagger}(\phi) =\sum_m e^{im\phi}\hat{\rho}_m$.
We then compare the phase-imprinted state with the original state through the overlap
\begin{equation} \label{eq:overlap}
    \mathcal{F}(\phi)
    =\operatorname{Tr}\!\left[\hat{\rho}\hat{\rho}(\phi)\right]
    =\sum_m I_m e^{im\phi}.
    \end{equation}
Here, $I_m=\operatorname{Tr}\!\left[\hat{\rho}_{-m}\hat{\rho}_m\right]$ is the MQC intensity of order~$m$. Thus, measuring $\mathcal{F}(\phi)$ as a function of the imprinted phase and Fourier transforming the result directly yields the complete distribution of MQC intensities.

Operationally, this measurement requires preparing the state, imprinting the collective phase, reversing the preparation, and measuring the probability of returning to the initial state. We implement this reversal through an adiabatic round trip~\cite{LewisSwan2020}, avoiding the need to reverse the sign of the full many-body Hamiltonian. The resulting echo protocol is closely related to measurements of out-of-time-ordered correlators used to probe entanglement and information scrambling~\cite{Gaerttner2017,Gaerttner2018}; further details are provided in the Methods and Extended Data Fig.~\ref{figure-adiabatic}. We now apply this framework to the extended toric code.

\begin{figure*}[t!!]
\centering
\includegraphics[width=0.85\linewidth]{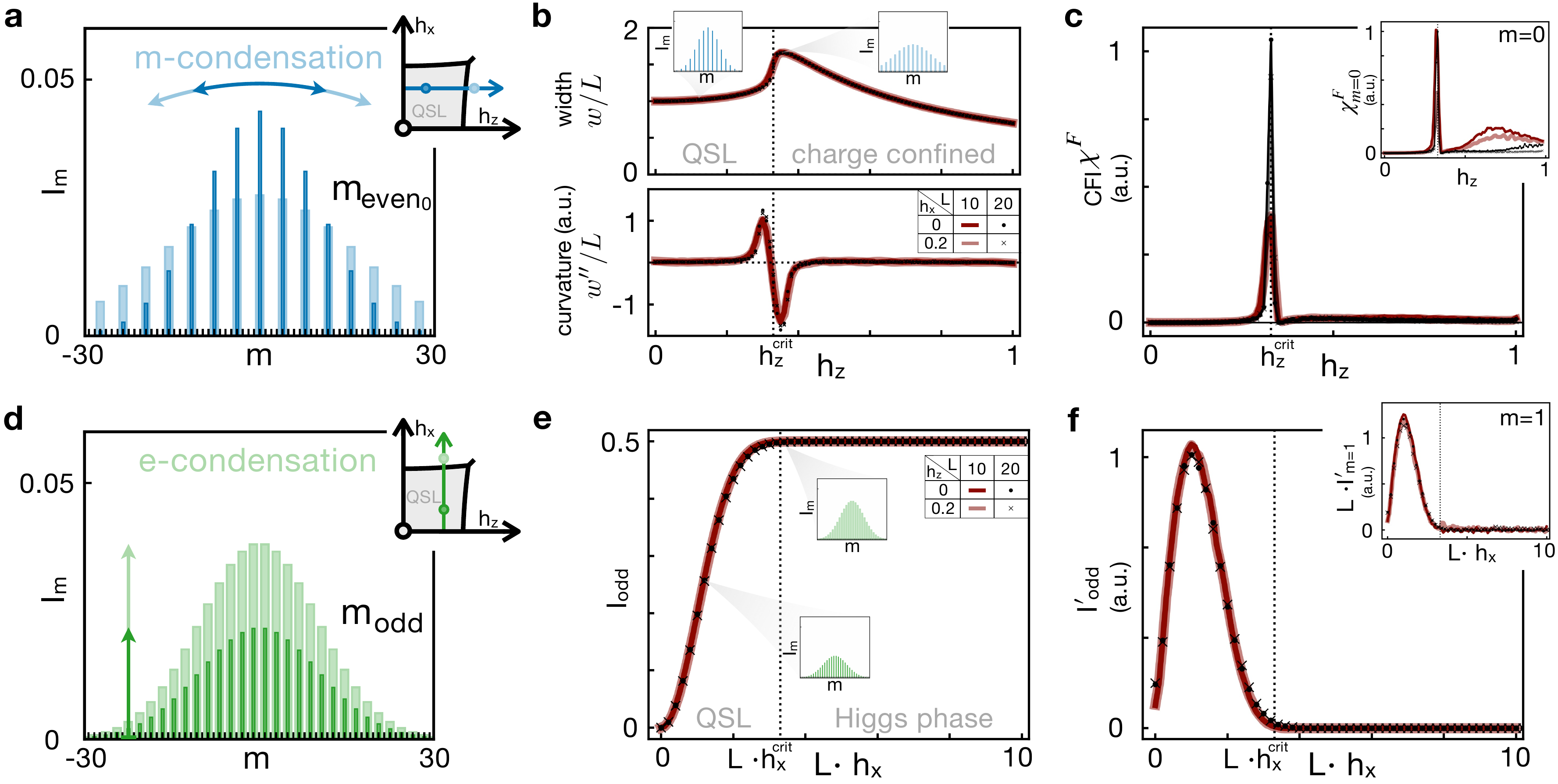}
\caption{\textbf{Extended toric code.}
\textbf{a} At the condensation transition the m-anyons strongly fluctuate leading to a response in all coherence sectors and critical broadening of the distributions, see also Fig.~\ref{figure-1}f. For illustration, we only plot the $m_{\rm even_0}$ sector for $h_z=0.13$ (dark blue) and $h_z=0.35$ (light blue) at~$h_x=0.2$; here~$L=10$.
\textbf{b} Top: The width~$w$ of the full distribution~$I_m$, after rescaling by system size~$L$. The inflection point coincides with the critical point, robust against finite size scaling and independent of the density of open strings (set by $h_x=0,0.2$). Bottom: The curvature $w'' = d^2w/dh_z^2$ distinguishes the QSL from charge-confined phase.
\textbf{c} The CFI shows strong response at the phase transition indicating critical broadening of the distribution. Inset: The CFI of the $m=0$~peak only.
\textbf{d} Across the Higgs transition, e-anyons (black circles) proliferate forming strong coherences in the odd-$m$ sector. For illustration, we only plot the $m_{\rm odd}$ sector for $h_x=0.13$ (dark green) and $h_x=0.35$ (light green) at~$h_z=0.2$; here~$L=10$.
\textbf{e} The weight of the odd coherences~$I_{\rm odd}$, after rescaling the field~$L \cdot h_x$.
\textbf{f} The fluctuations~$I'_{\rm odd}$ distinguish the QSL from Higgs phase. Inset: The fluctuations evaluated only at the $m=1$ peak.
Observables involving derivatives are estimated from numerical data using a filter routine to mitigate noise, hence given in arbitrary units (a.u.), see Methods.}
\label{figure-2}
\end{figure*}

\mysection{III. Connecting MQCs to quantum spin liquids}
We now turn to the paradigmatic extended toric code shown in Fig.~\ref{figure-1}c. The (dimensionless) Hamiltonian is given by
\begin{align}
\begin{split} \label{eq:tc-hamiltonain}
    \hat{H} = &-\sum_{+}\hat{A}_{+} -\sum_{\Box}\hat{B}_{\Box}
    - h_z\sum_{j}\hat{\sigma}^z_j - h_x\sum_{j}\hat{\sigma}^x_j,
\end{split}
\end{align}
where~$\hat{\sigma}^{x,z}_j$ describe Pauli spins residing on the links of a square lattice with~$L\times L$ plaquettes and periodic boundaries.
The first two terms are the toric-code star~$\hat{A}_+ = \prod_{j \in +}\hat{\sigma}^z_j$ and plaquette term~$\hat{B}_{\Box}=\prod_{j \in \Box}\hat{\sigma}^x_j$ summed over all vertices~$+$ and plaquettes~$\Box$, respectively. The~$h_z$ and~$h_x$ term penalize electric strings and create open strings, respectively.
Unless otherwise specified, we compute the phase diagram of Eq.~\eqref{eq:tc-hamiltonain} using worldline Quantum Monte Carlo (QMC) in $L=10,20$ systems~\cite{Linsel2026,Linsel2026a} at inverse temperatures~$\beta=L$. This temperature is sufficiently below the energy gap, such that snapshots sample the ground state. We obtain the MQC intensities~$I_m$ from these snapshots, see Methods.

For~$h_x=h_z=0$, the model is an exactly solvable \Ztwo~gapped QSL described by the toric code. In this limit, the star term imposes a closed-loop constraint (Gauss' law), such that the ground state consists of only closed-loop configurations resonantly fluctuating under the action of the plaquette term. 
The state can be nicely visualized through the corresponding pure density matrix,
\begin{align}
    \hat{\rho}_{\rm QSL}= \cdots +\ket{\raisebox{-0.2em}{\includegraphics[height=1em]{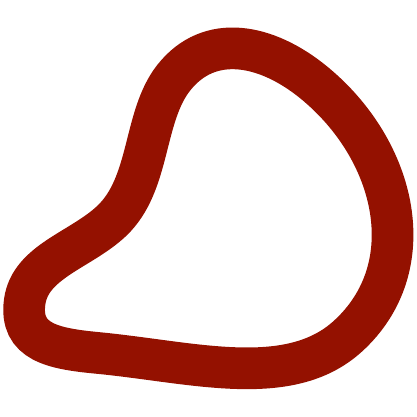}}}\!\!\bra{\raisebox{-0.2em}{\includegraphics[height=1em]{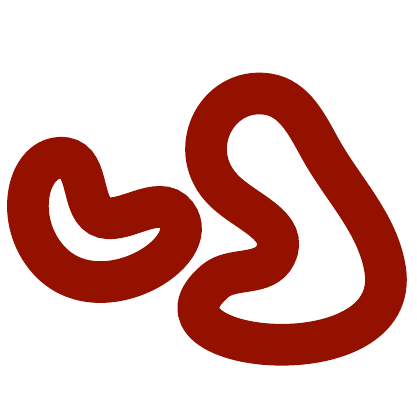}}} +\ket{\raisebox{-0.2em}{\includegraphics[height=1em]{icons/loop1.pdf}}}\!\!\bra{\raisebox{-0.2em}{\includegraphics[height=1em]{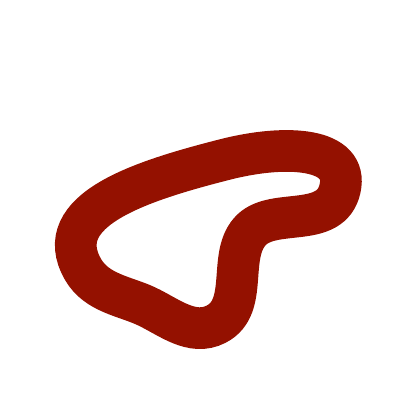}}} + \cdots,
\end{align}
whose off-diagonal elements encode coherences between distinct closed-loop configurations contributing to the weight in the MQC spectrum.

Away from the exactly solvable point, the perturbation~$\propto h_z$ acts as a string tension by imposing an energy cost on electric field lines. Conversely, the perturbation~$\propto h_x$ flips individual electric fields, thereby creating open strings of flipped spins whose endpoints correspond to electric e-anyons (violations of Gauss' law at vertices), also called matter excitations in gauge theory language. Similarly, excitations of the plaquette term, generated by the term~$\propto h_z$, correspond to magnetic m-anyons, see Fig.~\ref{figure-1}a. Upon increasing either field beyond a critical value~$h_{x,z}^{\rm crit} \approx 0.33$, the system undergoes a quantum phase transition into a topologically trivial phase~\cite{Trebst2007,Wu2012}, see Fig.~\ref{figure-1}c. This transition can be understood as the condensation of e- or m-anyons, respectively. Since individual anyons are created only at the endpoints of non-local string operators, the transition is not characterized by a conventional local order parameter.

This raises the question whether the intensities~$I_m$ contain sufficient information to distinguish different phases, as they naturally combine both global and local information.
To build intuition, we classify the coherence sectors according to the parity of spin flips~$m$ connecting the underlying configurations, illustrated in Fig.~\ref{figure-1}d. 
This separates the coherences into three classes:
(i) $m_{\mathrm{even}_0} \in 4\mathbb{Z}$, dominated by plaquette fluctuations,
(ii) $m_{\mathrm{even}_2} \in 4\mathbb{Z}+2$, dominated by string fluctuations, and
(iii) $m_{\mathrm{odd}} \in 2\mathbb{Z}+1$, associated with open strings or matter (e-anyon) fluctuations.
This characterization follows from the plaquette~$\hat{B}_{\square}$ and spin flip~$\hat{\sigma}^x$ processes generated by the toric-code Hamiltonian, and provides the physical interpretation of the coherence sectors throughout this work.

Before we quantitatively discuss the phase transitions, we establish the reference coherence spectrum of the toric code, see Fig.~\ref{figure-1}e.
Since the ground state consists exclusively of closed loops, odd coherence sectors vanish, $I_{m_{\rm odd}} \equiv 0$.
The remaining sectors $I_{m_{\mathrm{even}_0}}$ and $I_{m_{\mathrm{even}_2}}$ follow Gaussian distributions with width $w_{\rm tc}=L$,
consistent with the central-limit scaling, $w \propto \sqrt{N_{\rm spins}}\propto L$ spins.
This reference provides the basis for identifying how the coherence spectrum reorganizes across the phase diagram.

Along the two axes of the phase diagram in Fig.~\ref{figure-1}c, the phases can conventionally be distinguished using Wilson and 't~Hooft loops. These non-local observables are constructed by multiplying spin operators along a closed path and diagnose the confinement of the two types of anyons. Their expectation values decay with the size of the loop: whether this decay is controlled by the length of its boundary --the perimeter law-- or by the area it encloses --the area law-- distinguishes the topological and trivial phases~\cite{Wegner1971,Kogut1979}. 

Away from the axes of the phase diagram, fluctuations generate open strings that obscure the simple distinction between area- and perimeter-law scaling~\cite{FradkinShenker1979}. More refined disorder parameters can diagnose confinement~\cite{Gregor2011,Fredenhagen1986,Semeghini2021,Verresen2021,Linsel2024}, but their interpretation depends on the measurement basis~\cite{Wang2025}. Basis-independent quantities such as the topological entanglement entropy~$\gamma$ provide a more intrinsic characterization, but remain challenging to access experimentally~\cite{Satzinger2021}. MQC spectroscopy directly probes the underlying many-body coherences and therefore provides complementary information. \newline

\noindent\textbf{Confinement transition.}

At the confinement transition, the m-anyons proliferate and eventually condense, destroying the QSL, see Fig.~\ref{figure-1}c. A single m-anyon cannot be created by changing a spin locally. Instead, a sequence of spin operations must be applied along a non-local path, creating a pair of isolated m-anyons at its two endpoints. As the transition is approached, strings of increasing length proliferate throughout the system. The condensation is therefore encoded in the coherence structure of these non-local strings rather than in a conventional local order parameter.

MQC spectroscopy is sensitive to m-anyon condensation through the collective phase twist $\hat{W}(\phi)$ defined in Eq.~\eqref{rot}. Although this rotation is generated by a sum of local spin operators, its expansion contains products of spin operations forming strings of arbitrary length. In the electric-field basis, an open string operator~$\prod_{j \in \mathcal{C}}\hat{\sigma}^z_j$ creates a pair of m-anyons at the endpoints of the path~$\mathcal{C}$. The response to the phase twist therefore samples the non-local string fluctuations that proliferate as the m-anyons condense.

This sensitivity appears in the MQC spectrum as a redistribution of spectral weight and a pronounced broadening near the transition, as shown in Figs.~\ref{figure-1}f and~\ref{figure-2}a. We quantify this effect through the width $w(h_z)$ of the full coherence distribution, normalized by the linear system size~$L$ in Fig.~\ref{figure-2}b (top). In the QSL regime, $w(h_z)/L$ approaches the toric-code value $w_{\rm tc}/L=1$. Near the transition, the spectrum broadens critically. After rescaling, the results for $L=10$ and $L=20$ collapse onto a common curve for both $h_x=0$ and $h_x=0.2$, demonstrating that the scaling is robust to the presence of matter excitations. Beyond the transition, the spectrum narrows again. Deep in the charge-confined phase, $h_z\gg 1$, the spins polarize along the $z$~direction and the coherence distribution collapses to a single peak at~$m=0$.

The transition is revealed particularly clearly by the curvature~$w''=d^2w/dh_z^2$, see Fig.~\ref{figure-2}b (bottom). As the system crosses from the QSL into the charge-confined phase, $w(h
_z)$ develops an inflection point at which $w''$ changes sign. The corresponding zero crossing therefore provides a sharp marker of the phase boundary.

The same redistribution is reflected in the classical Fisher information (CFI)~\cite{Gaerttner2018}:
\begin{align}
    \chi^F(h_z) = \sum_m\chi^F_m(h_z) = \sum_m I_m \left(\frac{\partial_{h_z}I_m}{I_m}\right)^2.
\end{align}
The CFI measures how sensitively the entire MQC distribution responds to a small change in $h_z$. It becomes large when a small parameter variation produces a substantial redistribution of coherence among the different sectors. In Fig.~\ref{figure-2}c, we plot the response~$\chi^F$ showing a pronounced peak at the critical point, establishing the CFI as a sensitive diagnostic of m-anyon condensation. Remarkably, the  experimentally accessible~$m=0$ sector already reproduces this behaviour as shown in the inset.

The MQC signatures remain robust when matter fluctuations are introduced through a finite~$h_x$. For the values of $h_x$ considered here, the critical broadening, scaling collapse and Fisher-information peak remain clearly visible. Conventional loop-based observables can instead become more difficult to interpret when open strings are present, particularly near the transition~\cite{Gregor2011,Linsel2024}. MQC spectroscopy therefore provides a complementary diagnostic of confinement that remains effective in the presence of matter fluctuations. 
\newline

\noindent\textbf{Higgs transition.}

We next turn to the Higgs transition, driven by increasing the spin-flip field~$h_x$, see Fig.~\ref{figure-1}c. Acting on a closed-loop configuration, a spin flip creates an open string whose two endpoints correspond to e-anyons. For weak~$h_x$, these excitations are gapped and occur only as dilute, short-range virtual pairs, and the system remains in the QSL phase. As $h_x$ increases, open strings proliferate and their endpoints become coherently delocalized across the system. In the thermodynamic limit, this proliferation signals the condensation of e-anyons and the onset of the Higgs phase.

The odd MQC sectors provide a sensitive probe of this process. Closed-loop configurations contribute only to even coherence orders, whereas odd orders arise from coherences involving configurations with open strings. The increasing weight of $I_{m_{\rm odd}}$ therefore tracks the growing coherence associated with e-anyon fluctuations. As shown by the representative spectra in Fig.~\ref{figure-2}d, this weight grows as the system approaches and crosses the Higgs transition.

Fig.~\ref{figure-2}e shows the weight of the total odd coherence sector, $I_{\rm odd}=\sum_{m_{\rm odd}}I_m$, for $L=10$ (red) and $L=20$ (black). After rescaling the field~$L \cdot h_x$, we find excellent data collapse onto a single curve, independent of m-anyon fluctuations set by $h_z=0, 0.2$. The rescaling corrects for the fact that in large systems small~$h_x$ already leads to a macroscopic occupation of odd coherence sectors.
Our analysis reveals how coherences between matter excitations develop as the field~$h_x$ is increased. In the QSL, $I_{\rm odd}$ monotonically increases until it reaches a plateau beginning at~$L \cdot h_x^{\rm crit}$.
In the plateau region, the coherence distribution approximately resembles the one of trivially polarized spins in the $x$~direction. 

To distinguish the fluctuating regime of $I_{\rm odd}$ in the QSL from the weakly fluctuating regime in the trivial phase, we analyze the derivative $I'_{\rm odd} = dI_{\rm odd}/d(Lh_x)$, shown in Fig.~\ref{figure-2}f. The quantity follows an order parameter-like behaviour in the QSL phase, independent of the electric field~$h_z$. 
The experimentally accessible~$m=1$ sector already reproduces this behaviour as shown in the inset.

While the coherence spectrum contains characteristic signatures across the e-anyon condensation transition, it does not exhibit sharp signatures like the ones we find for m-anyon condensation, making a clear experimental measurement of the critical point challenging. We understand this behaviour through the choice of basis for the phase rotation~$W(\phi)$: The phase rotation generated by collective~$\hat{S}^z$ is an off-diagonal operation with respect to m-anyons, measuring the system's response to create anyon pairs. In contrast, the phase rotation only measures the already existing coherence between e-anyons, but does not create new pairs of e-anyons. By choosing the basis of phase rotation to have overlap with both anyon creation operators, the coherence distribution is expected to exhibit sharp signatures across both transitions~\cite{Linsel2026_prxq}.
\newline

\begin{figure}[t!!]
\centering
\includegraphics[width=\linewidth]{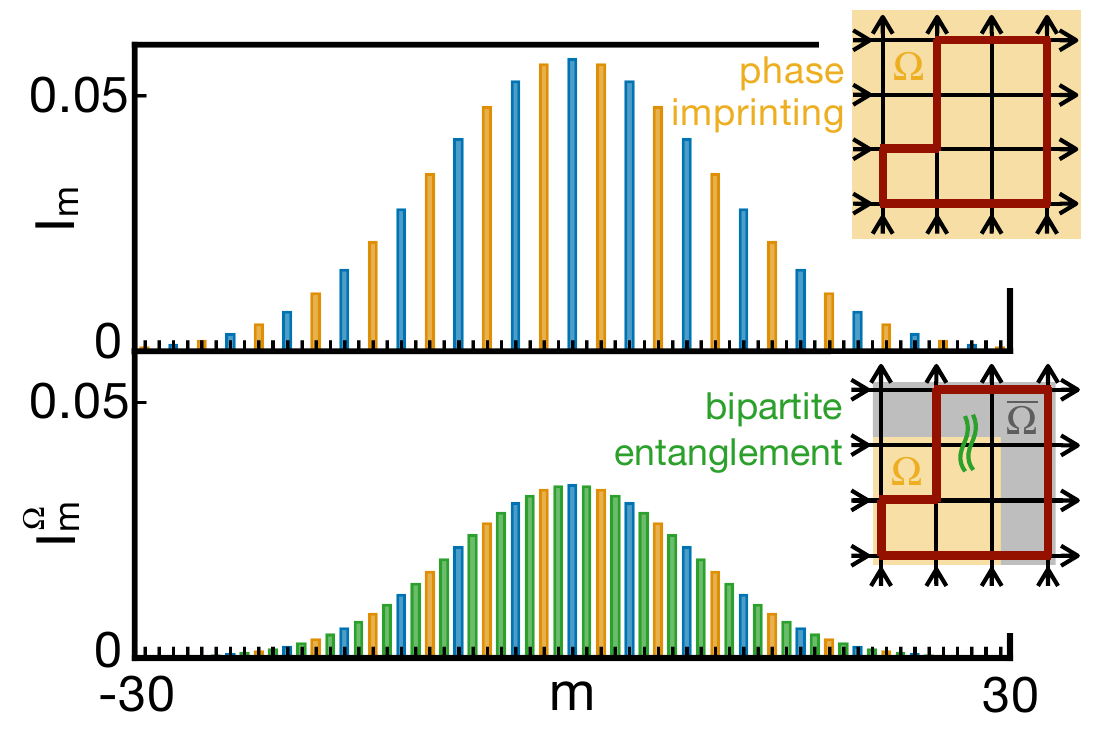}
\caption{\textbf{Bipartite entanglement.} 
We apply the phase imprinting on subsystems~$\Omega$ of size~$\ell \times \ell$ (orange shade), while the complement~$\bar{\Omega}$ does not acquire a phase, resulting in MQC intensities~$I_m^\Omega$ that encode information shared across the bipartition. 
Top: If $\ell=L$ we recover the MQC distribution with weights only at even~$m$. Bottom: When applying the phase rotation only on subsystem~$\Omega$ with~$\ell < L$, intensities~$I_m^\Omega$ associated with odd number of spin flips~$m$ appear in the distribution, despite a global closed-loop constraint. Non-zero weight~$I_{\rm odd}^\Omega >0$ (green) is a witness for bipartite entanglement~$\Omega|\bar{\Omega}$. 
}
\label{figure-3}
\end{figure}

\begin{figure*}[t!!]
\centering
\includegraphics[width=\linewidth]{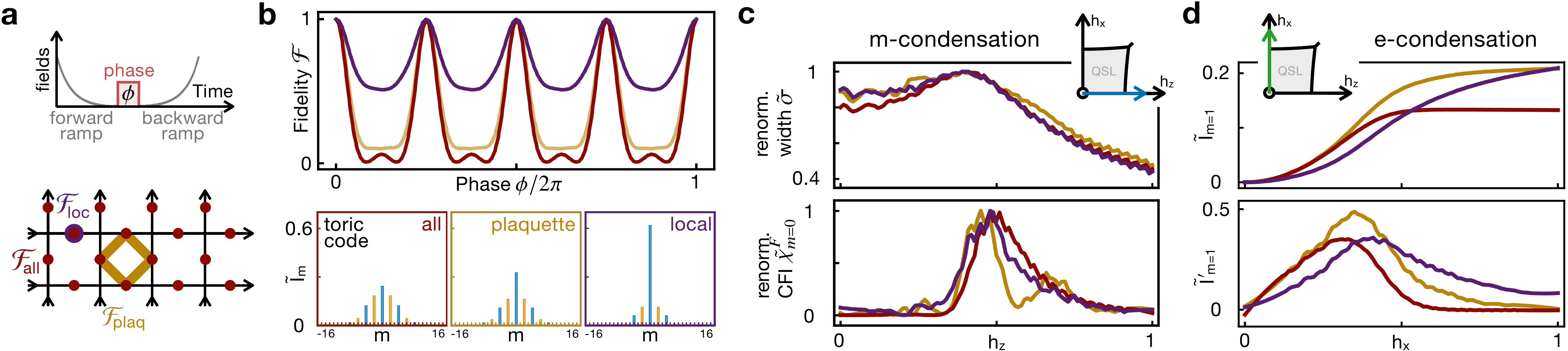}
\caption{\textbf{Experimental protocol and local fidelities.} \textbf{a} Top: Adiabatic round trip protocols, combined with a phase imprinting~$\phi$ on the many-body state~\cite{Karch2026}, provide an experimentally accessible MQC spectroscopy. Bottom: We simulate the exact dynamics of the round-trip protocol in a small system. The return fidelity to the initial product state is measured globally~$\mathcal{F}_{\rm all}$ (red), and on subsystems with four spins~$\mathcal{F}_{\rm plaq}$ (yellow) and one spin~$\mathcal{F}_{\rm loc}$ (purple). \textbf{b} Top: We plot the return fidelities as we vary the phase imprinting~$\phi$ at the toric-code point. Bottom: The corresponding~$\tilde{I}_m$ distributions obtained by a Fourier transformation of~$\mathcal{F}(\phi)$; the notation $\tilde{I}_m$ highlights that local fidelities acquire only partial information of the full~$I_m$, see Methods. \textbf{c} We analyze the~$\tilde{I}_m$ for the m-condensation crossover ($h_x=0$). Top: The normalized width~$\tilde{\sigma}$ obtained from the different subsystem fidelities. Bottom: The CFI from the~$m=0$ peak shows sharp features, retained in all local fidelity measurements. \textbf{d} Top: We analyze the~$\tilde{I}_{m=1}$ peak for the e-condensation crossover ($h_z=0$). Bottom: The derivative~$\tilde{I}'_{m=1} = d\tilde{I}_{m=1}/dh_x$. While qualitative features agree, the point at which the plateau is reached deviates between global and local fidelities.    }
\label{figure-4}
\end{figure*}

\noindent\textbf{Subsystems \& entanglement witness.}


We next probe a subsystem~$\Omega$ by applying a phase rotation~$\hat{W}_\Omega(\phi) = e^{i\phi \hat{S}^z_\Omega}$ with~$\hat{S}^z_\Omega=
\frac{1}{2}\sum_{j \in \Omega}\hat{\sigma}^z_j$, see Fig.~\ref{figure-3}a. The extracted intensities~$I_m^\Omega$, analogous to Eq.~\eqref{eq:overlap}, retain information about the coherence between~$\Omega$ and its complement~$\bar{\Omega}$.
For closed loops, $h_x=0$, the intensities~$I_{m_{\rm odd}}^\Omega$ provide a bipartite entanglement witness:
\begin{align*}
    I_{m_{\rm odd}}^\Omega > 0~\cap \text{closed loops} \Rightarrow \text{bipartite entanglement } \Omega | \bar{\Omega}
\end{align*}

To show this, we consider subsystems of size~$\Omega=\ell \times \ell$ on a square lattice with an even number of plaquettes, and organize the configurations by even and odd number of flipped spins in the subsystems. 
If the system were separable in~$\Omega$ and~$\bar{\Omega}$, it reads $\ket{\psi_{\rm sep}} = (a\ket{{\rm even},\Omega}+b\ket{{\rm odd},\Omega}) (c\ket{{\rm even},\bar{\Omega}}+d\ket{{\rm odd},\bar{\Omega}})$.
The global closed loop constraint restricts the parity of total flipped spins to be even.
However, this is violated by a generic separable state because it contains contributions $\ket{{\rm even},\Omega}\ket{{\rm odd},\bar{\Omega}}$ or $\ket{{\rm odd},\Omega}\ket{{\rm even},\bar{\Omega}}$, which are not allowed.
Therefore, the only two separable states consistent with the closed loop constraint are $\ket{\psi_{\rm sep}} =\ket{{\rm even},\Omega}\ket{{\rm even},\bar{\Omega}}$ or $\ket{\psi_{\rm sep}} =\ket{{\rm odd},\Omega}\ket{{\rm odd},\bar{\Omega}}$.
Those states always have subsystem $I_m^\Omega$ at even~$m$, see Extended Data Fig.~\ref{figure-bipartite}. The argument extends to general separable mixed states.
Consequently, a closed-loop state with non-zero $I_{m_{\rm odd}}^\Omega>0$ has bipartite entanglement between~$\Omega$ and~$\bar{\Omega}$.

Fig.~\ref{figure-3} shows the qualitatively different distributions of the global intensities~$I_m$ and subsystem intensities~$I^\Omega_m$.
While non-zero even~$m$ sectors in the global protocol reveal coherences between closed-loop configurations, the appearance of odd~$m$ sectors in the subsystem protocol detects that the closed loops are coherently shared across a bipartition, providing an experimentally accessible witness for entanglement.

\textit{Quantum spin lakes.}
The subsystem coherences~$I_{m_{\rm odd}}^\Omega$ spatially distinguish a classical closed-loop gas from a coherently entangled loop gas.
This suggests a possible probe for quantum spin lakes~\cite{Sahay2022}, which are finite regions of dynamically prepared QSLs with coherence length~$\xi$~\cite{Semeghini2021,Giudici2022,Mauron2025,Karch2025}.
We consider an ansatz~$\hat{\rho}_{\rm lake} = \sum_k p_k \hat{\rho}_k^{\rm QSL}\otimes \hat{\rho}_k^{\rm \overline{QSL}}$, where $\hat{\rho}_k^{\rm QSL}$ describes a coherent QSL of extent~$\xi$ and $\hat{\rho}_k^{\rm \overline{QSL}}$ its trivial complement. The two regions are separable by construction. 
To probe~$\xi$, we suggest to vary the linear size~$\ell$ of the subsystem. For $\ell \lesssim \xi$, the subsystem is within the coherent loop region and exhibits~$I_{\rm odd}^\Omega>0$.
Once~$\ell$ exceeds the coherent region, the bipartition crosses the separable boundary and the entanglement witness vanishes. The subsystem dependence of $I_{\rm odd}^\Omega$ can therefore provide a probe of the spin-lake coherence length.

\mysection{IV. Experimental implementation}
In an experiment, the direct measurement of~$\mathcal{F}(\phi)$ in Eq.~\eqref{eq:overlap} is difficult as it requires measuring an overlap between many-body wavefunctions.
This can be overcome through time-reversal protocols or adiabatic round trips~\cite{LewisSwan2020,Karch2026} connecting the many-body to a simple product state, which has been demonstrated in trapped ion experiments~\cite{Gaerttner2017} and which provides an alternative interpretation of MQCs as out-of-time-ordered correlators~\cite{Gaerttner2018}.

Since~$\hat{\rho} =\hat{\rho}_{\rm QSL}$ is the density matrix of a ground-state wavefunction, the fidelity~$\mathcal{F}(\phi)$ therefore corresponds to the population remaining in the ground state after a phase twist induced by~$\hat{W}(\phi)$.
Experimentally, the many-body state~$\hat{\rho}_{\rm QSL}$ can be prepared via an adiabatic protocol connecting the ground state of a simple, unentangled system to the many-body ground state in a forward ramp, see Fig.~\ref{figure-4}a.
Applying the global phase twist~$\hat{W}(\phi)$ and reversing the protocol by a backward ramp~\cite{LewisSwan2020}, connects the many-body ground state back to the simple product state, such that the return fidelity equals~$\mathcal{F}(\phi)$, see Methods and Extended Data Fig.~\ref{figure-adiabatic}.

Of experimental relevance is not only the return fidelity measured on all spins~$\mathcal{F}_{\rm all}(\phi)$, but also the more noise-resilient fidelity measurement on subsystems~$\mathcal{F}_{\rm sub}(\phi)$~\cite{Karch2025}.
However, the information remaining within the subsystem fidelity measurement becomes a ramp-dependent problem, including the adiabaticity and shape of the ramp.
To demonstrate that crucial information can be retained in the subsystem fidelities, we simulate an adiabatic roundtrip using exact diagonalization on a small system, see Fig.~\ref{figure-4}a.
As we reduce the subsystem size, high frequency oscillations in~$\mathcal{F}_{\rm sub}(\phi)$ (corresponding to large~$m$) are suppressed relative to~$\mathcal{F}_{\rm all}(\phi)$, see Fig.~\ref{figure-4}b.
Nonetheless, the reconstructed intensities~$\tilde{I}_m$ capture the qualitative features of the toric code's coherences.

The confinement transition, shown in Fig.~\ref{figure-4}c, is detected through the adiabatic protocol even by measuring the local return fidelity of a single spin.
The Higgs transition, shown in Fig.~\ref{figure-4}d, is well captured by the full system fidelity~$\mathcal{F}_{\rm all}(\phi)$.
While qualitative features persist in subsystem fidelities, the quantitative measurement of the critical point is not reproduced.
This underscores again that a full characterization of the phase diagram requires phase twists that probe both the response of e- and m-anyons~\cite{Linsel2026_prxq}.

\mysection{V. Outlook}

We have shown that the MQC spectrum provides more than a measure of the total amount of coherence in a many-body state. Its structure can be connected directly to the anyonic excitations of a gapped QSL and reveals how these excitations condense across the surrounding phase diagram. At the same time, the subsystem MQCs can provide a direct witness for bipartite entanglement. Together with the local and experimentally accessible protocols developed here, these results establish MQCs as complementary probes of both the excitations and non-local organization of topological quantum matter. More broadly, they illustrate how the central strategy of quantum sensing -- converting otherwise hidden coherence into a simple measurable signal -- can be extended to strongly correlated many-body systems.

An immediate direction is to apply this framework to states prepared dynamically in quantum simulators. 
Extending MQC measurements to non-adiabatic protocols~\cite{Sahay2022,Giudici2022,Mauron2025} would enable direct comparisons with recent experiments on gapless U(1) QSL candidates~\cite{Geim2026,Bornet2026,Karch2026}.
The bipartite entanglement witness accessible through subsystem MQC may therefore help distinguish a mesoscopic, dynamically generated spin lake from a thermodynamic spin-liquid phase. 

More generally, extending the framework beyond the toric code will require identifying collective phase-imprinting operations that couple naturally to the relevant emergent excitations. This could make MQCs useful for addressing long-standing questions concerning candidate QSLs in models such as the kagome Heisenberg antiferromagnet and the square-lattice $J_1-J_2$
model~\cite{Meng2026}, as well as for exploring topological matter in quantum simulators and solid-state systems~\cite{Halimeh2025}. Similar ideas applied to Hubbard systems could probe the many-body coherence associated with unconventional pairing, opening a route from coherence-based diagnostics of spin liquids to broader classes of strongly correlated quantum matter.

\mysection{Acknowledgements}
We thank Monika Aidelsburger, Antoine Browaeys, Raphael Kaubruegger, and Simon Karch for insightful discussions.
LH and AMR acknowledge support by the Simons Collaboration on Ultra-Quantum Matter, which is a grant from the Simons Foundation (651440).
This project is also supported by the NSF JILA-PFC PHY-2317149 and the US Department of Energy, Office of Science, National Quantum Information Science Research Centers, Quantum Systems Accelerator.
S.M.L. acknowledges funding from the European Research Council (ERC) under the European Union’s Horizon 2020 research and innovation program -- ERC Starting Grant SimUcQuam (Grant Agreement No. 948141) and QuantERA II (Grand Agreement No. 101017733), by the QuantERA grant DYNAMITE, and by the Deutsche Forschungsgemeinschaft (DFG, German Research Foundation) under project number 499183856 and under Germany's Excellence Strategy -- EXC-2111 -- project number 390814868.
L.P. acknowledges financial support from the Deutsche Forschungsgemeinschaft (DFG), project Nr 530111096. The project/research is part of the Munich Quantum Valley, which is supported by the Bavarian state government with funds from the Hightech Agenda Bayern Plus.\\


%

\pagebreak


\begin{center} 
\textbf{\large Methods}
\end{center}

\setcounter{figure}{0}
\renewcommand{\figurename}{Extended Data Figure}
\renewcommand{\thefigure}{\arabic{figure}}

\subsection*{Multiple quantum coherences from adiabatic round trips}

\begin{figure}[t!!]
\centering
\includegraphics[width=0.8\linewidth]{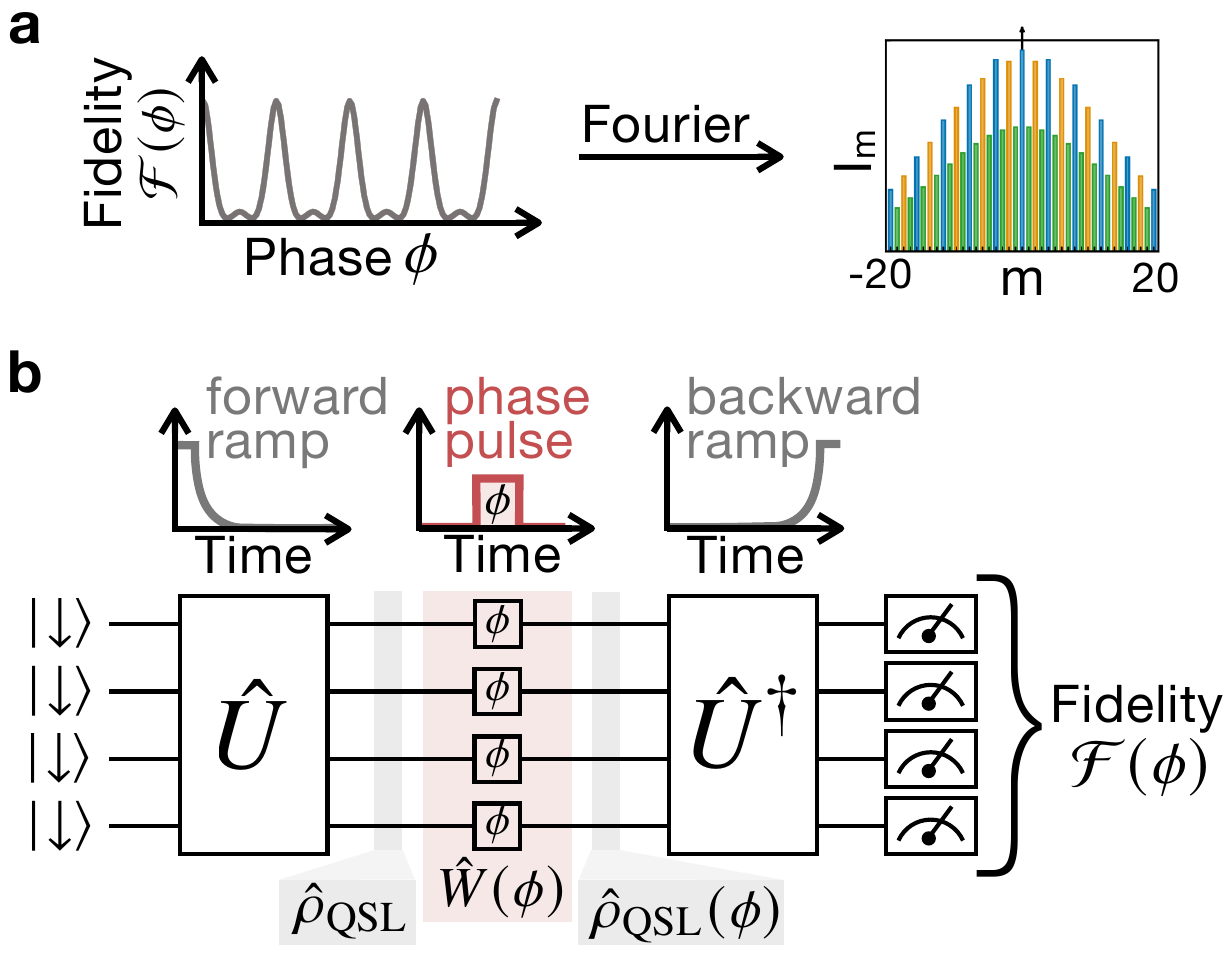}
\caption{\textbf{Fourier spectroscopy and adiabatic round trip.} 
\textbf{a} The fidelity~$\mathcal{F}(\phi)$ probes the response of the quantum state~$\hat{\rho}_{\rm QSL}$ to a collective phase twist~$\hat{W}(\phi)$. This provides a spectroscopic tool to access the coherences~$I_m$, which are the Fourier amplitudes of~$\mathcal{F}(\phi)$. The exemplary coherence distribution corresponds to $h_x=0.2$ and $h_z=0.5$ for $L=10$ in the extended toric code. The distribution illustrates that the three sectors $m_{\rm even_0}$ (blue), $m_{\rm even_2}$ (orange) and $m_{\rm odd}$ (green) are captured by three separate Gaussian envelopes.
\textbf{b} The fidelity~$\mathcal{F}(\phi)$ of a many-body state is experimentally accessible through an adiabatic round trip protocol effectively realizing time reversal.   }
\label{figure-adiabatic}
\end{figure}

In the main text, we established the MQC framework, relating the intensities~$I_m$ to the overlap~$\mathcal{F}(\phi)=\operatorname{Tr[\hat{\rho}\hat{\rho}(\phi)]}$, see Extended Data Fig.~\ref{figure-adiabatic}a.
For the ground states we consider the overlap has a simple physical interpretation: It measures how much population remains in the ground state after applying a phase rotation~$\hat{W}(\phi)$.
While this many-body overlap is very challenging to obtain experimentally, it becomes accessible through an adiabatic protocol by connecting the ground-state population to a simpler system, as we explain in the following.

Inspired by time-reversal protocols studied in trapped ions, we propose a natural scheme applicable for analog quantum simulators of QSLs through forward and backward propagation of an adiabatic ramp~$h(t)$~\cite{LewisSwan2020}, see Extended Data Fig.~\ref{figure-adiabatic}b. Starting from an initial product state~$\hat{\rho}_{\rm init}$, that is the unique ground state of a non-interacting Hamiltonian~$\hat{Z}$, the system adiabatically follows the ground state of a slowly varying Hamiltonian~$\hat{H}(t)=\hat{H} + h(t)\hat{Z}$ connecting the initial state to the target many-body ground state of~$\hat{H}$ via the unitary~$\hat{U}$. After imprinting the phase~$\hat{W}(\phi)$, the Hamiltonian is ramped backwards by reversing the direction of~$h(t)$ but \textit{without} inverting the sign of the Hamiltonian, effectively described via the unitary~$\hat{U}^\dagger$ in the adiabatic limit. While the phase imprinting creates excitations in~$\hat{\rho}$, the backward ramp maps the remaining ground-state population of~$\hat{H}$ to the ground-state population of~$\hat{Z}$. 

As a result, the overlap of the initial and final state,~$\mathcal{F}(\phi)=|\langle \psi_{\rm final}(\phi) |\psi_{\rm init}\rangle|^2$, is directly related to the coherences~$I_m$ in Eq.~\eqref{eq:overlap}. Since~$|\psi_{\rm init}\rangle$ corresponds to a simple product state, this fidelity is accessible by projective measurements on the final state. 
This can be seen by defining the initial (final) state density matrix~$\hat{\rho}_{\rm init}$ ($\hat{\rho}_{\rm final}$). The initial state is related to the QSL by the adiabatic ramp~$\hat{\rho}_{\rm QSL}=\hat{U}\hat{\rho}_{\rm init}\hat{U}^\dagger$.
The overlap after the adiabatic round trip is then given by
\begin{align}
    \begin{split} \label{eq:adiabatic-roundtrip}
        \mathrm{Tr}[\hat{\rho}_{\rm final}\hat{\rho}_{\rm init}] &= \mathrm{Tr}[\hat{U}^\dagger\hat{W}(\phi)\hat{U}\hat{\rho}_{\rm init}\hat{U}^\dagger\hat{W}^\dagger(\phi)\hat{U}\hat{\rho}_{\rm init}] \\
        &=\mathrm{Tr}[\hat{W}(\phi)\hat{U}\hat{\rho}_{\rm init}\hat{U}^\dagger\hat{W}^\dagger(\phi)\hat{U}\hat{\rho}_{\rm init}\hat{U}^\dagger] \\
        &=\mathrm{Tr}[\hat{\rho}_{\rm QSL}(\phi)\hat{\rho}_{\rm QSL}].
    \end{split}
\end{align}
We emphasize that the adiabatic forward and backward ramp effectively realizes time reversal without the requirement of changing the sign of the Hamiltonian~\cite{LewisSwan2020}. Furthermore, this relates MQCs to out-of-time-ordered correlators~\cite{Gaerttner2017,Gaerttner2018}.

\subsection*{Multiple quantum coherences from ground-state snapshots}
  
We show how the distributions~$I_m$ can be computed from ground-state snapshots. To this end, we label the Fock states~$\ket{M, n_M}$ by their total magnetization~$M$ and an additional index~$n_M=1,...,\mathcal{N}_M$ with the multiplicity~$\mathcal{N}_M = {N_{\rm spins} \choose M +N_{\rm spins}/2}$. Therefore, we express the wavefunction of spin-$1/2$ particles by
\begin{align}
    \ket{\psi} = \sum_M \sum_{n_M} \alpha_{M,n_M} \ket{M, n_M},
\end{align}
with a corresponding pure density matrix~$\hat{\rho} = \sum_m \hat{\rho}_m$, where we have organized sectors in the density matrix, whose total magnetization differs by~$m$:
\begin{align}
    \hat{\rho}_m &= \sum_{m=M-M'} \sum_{n_M} \sum_{n_{M'}} \alpha_{M,n_M} \alpha^*_{M',n_{M'}} \ket{M, n_M} \bra{M', n_{M'}}
\end{align}
The coherences~$I_m$ correspond to the Frobenius norm of such sectors,
\begin{align} \label{eq:Im_QMC}
\begin{split}
    I_m = || \hat{\rho}_m ||^2_F &= \sum_{m=M-M'} \sum_{n_M} \sum_{n_{M'}} |\alpha_{M,n_M}|^2 |\alpha_{M',n_{M'}}|^2 \\
    &=\sum_{m=M-M'} p_M p_{M'} .
\end{split}
\end{align}
Conveniently, the probabilities~$p_M=\sum_{n_M}|\alpha_{M,n_M}|^2$ can be sampled efficiently from snapshots assuming \textit{pure} quantum states in our numerical simulations.

\subsection*{Quantum Monte Carlo}
We perform QMC simulations of the extended toric code using the \textsc{ParaToric} package~\cite{Linsel2026,Linsel2026a}. We simulate system sizes~$L=10, 20$ with periodic boundaries ($N_{\rm spins}=2L^2$) at inverse temperatures~$\beta = L$, such that the temperature is well below the energy gap, and the snapshots predominantly sample the ground state. The parameter sweeps of the extended toric code we consider are (i) $h_z=0$, $h_x=0,\ldots,1$, (ii) $h_z=0.2$, $h_x=0,\ldots,1$, (iii) $h_x=0$, $h_z=0,\ldots,1$, and (iv) $h_x=0.2$, $h_z=0,\ldots,1$, where the varying field is sampled on $101$~parameter points. For each parameter point, we obtain $10^5$~spatially resolved snapshots in the computational basis~$\hat{\sigma}^z$, allowing us to resolve even large~$|m|$ sectors in the QSL.

\subsubsection*{Numerical derivatives}

The observables we consider at the phase transitions, see Fig.~\ref{figure-2} and~\ref{figure-4}, involve first and second derivatives.
Despite our small spacing of data in the QMC simulations~$\Delta h_{x,z}=0.01$ obtaining accurate values of derivatives on the finite grid is noisy due to the small denominator of the approximate derivative.
In practice, this noise obscures features, such as the inflection point (Fig.~\ref{figure-2}b) or the monotonic behaviour (Fig.~\ref{figure-2}f), visible in the raw data.
To this end, we estimate the derivatives using a standard Savitzky–Golay filter routine.
Close to the critical point, where the derivatives have large magnitude, the precise numerical value of the derivative depends on the parameters of the filter routine.
Importantly, the qualitative features, such as zero crossing or peak positions, remain robust.

\begin{figure}[t!!]
\centering
\includegraphics[width=0.8\linewidth]{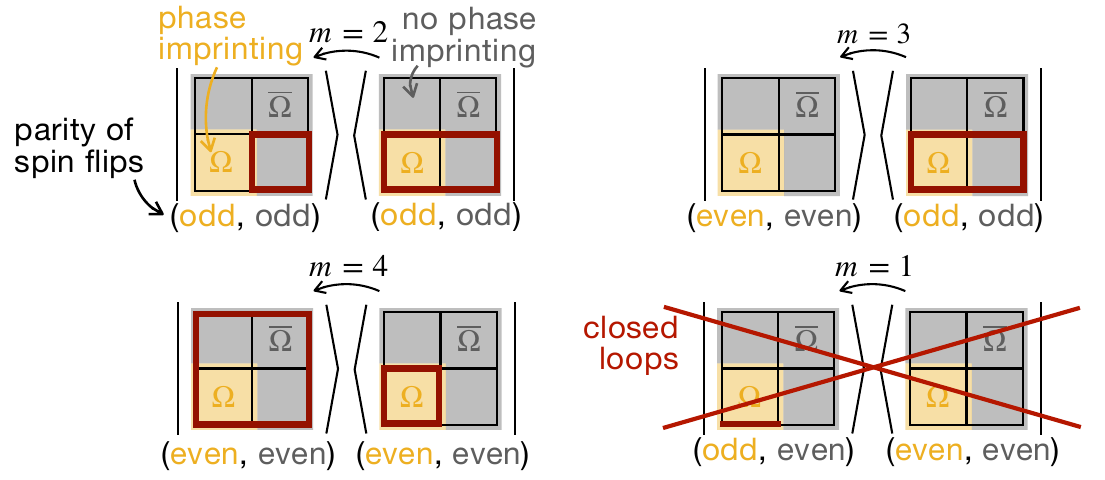}
\caption{\textbf{Subsystem coherences.} 
Subsystems~$\Omega$ (orange shade) acquire a phase, while the complement~$\bar{\Omega}$ does not acquire a phase. We illustrate exemplary contributions to the MQC intensities, where the emergent electric fields (red lines) correspond to flipped spins. We organize the coherence sectors by the parity of flipped spins in~($\Omega$, $\bar{\Omega}$) of each configuration. The closed loop constraint requires that the global parity is even.
  }
\label{figure-bipartite}
\end{figure}

\subsection*{Bipartite entanglement}

The global closed loop constraint requires the total parity of flipped spins to be even. Therefore, when we partition the system into regions~$\Omega$ and~$\bar{\Omega}$, we can analogously assign a parity of flipped spins within each subsystem. The global constraint enforces the two subsystem parities to be equal, hence the only allowed configurations have (even, even) and (odd, odd) parity on ($\Omega$, $\bar{\Omega}$), see Extended Data Fig.~\ref{figure-bipartite}.

Because in the subsystem measurement only the spins in region~$\Omega$ acquire a phase, contributions to $I_m^\Omega$ at even~$m$ and odd~$m$ can be understood from the coherence between different subsystem parity sectors. Crucially, intensities at odd~$m$ must originate from coherence between (even, even) $\leftrightarrow$ (odd, odd). In a state separable across the two subsystems, however, those coherences are not allowed under the global closed loop constraint.

\subsection*{Subsystem fidelities}
The MQC protocol directly relates the round trip fidelity to the coherence distribution~$I_m$, see Eq.~\eqref{eq:adiabatic-roundtrip}, by measuring the \textit{global} return fidelity~$\mathcal{F}_{\rm all}$ to the initial product state, see Fig.~\ref{figure-4}a.
In the presence of experimental white noise, the global fidelity~$\mathcal{F}_{\rm all}$ decays exponentially with system size, strongly limiting the application of the protocol in large systems on analog quantum simulators. To mitigate this effect, we instead consider local fidelities~\cite{Karch2025} and analyze the information retained in the oscillating signal~$\mathcal{F}_{\rm sub}(\phi)$.

To do so, we first discuss the physical quantity corresponding to local fidelities.
Recall that in our protocol the phase imprinting~$\hat{W}(\phi)$ transfers population from the many-body ground state into excited states, leading to the fidelity response~$\mathcal{F}(\phi)$. For an ideal adiabatic round-trip protocol, the many-body ground state is mapped onto a ground state of a simple system, i.e., a product state, such that a global fidelity measurement directly reproduces the full coherence signal, $\mathcal{F}_{\rm all}(\phi) = \mathcal{F}(\phi)$. 

In contrast, a local fidelity generally cannot discriminate between the ground state and excited states that share the same local reduced density matrix.
Consequently, the measured subsystem response~$\mathcal{F}_{\rm sub}(\phi)$ becomes sensitive to the details of the ramp protocol as it depends on how excited states are connected to the final system.
More specifically, for a measurement on a subsystem~$\Omega$ the local return fidelity is determined by the reduced density matrix~$\hat{\rho}_\Omega = \mathrm{Tr}_{\bar{\Omega}}\ket{\psi }\bra{\psi}$ obtained after tracing out the complement~$\bar{\Omega}$. The corresponding local projector onto the initial state is~$\hat{\mathcal{P}}_{\rm sub} = \mathbbm{1}_{\bar{\Omega}} \otimes \mathrm{Tr}_{\bar{\Omega}}\ket{\psi_{\rm init}}\bra{\psi_{\rm init}}$. 
In the adiabatic roundtrip protocol, this projector evolves under the time-dependent ramp as~$\hat{U}\hat{\mathcal{P}}_{\rm sub}\hat{U}^\dagger$.
Therefore, the local fidelity measurement can be expressed as
\begin{align}
    \mathcal{F}_{\rm sub}(\phi) = \mathrm{Tr}[\hat{\rho}(\phi) \hat{U}\hat{\mathcal{P}}_{\rm sub}\hat{U}^\dagger ] =\sum_m \tilde{I}_m e^{i\phi m},
\end{align}
where~$\hat{\rho}$ is the many-body state we want to probe. We have defined the approximate coherence distribution~$\tilde{I}_m$. The proximity of this distribution to the coherences~$I_m$ is thus determined by the operator spread of $\hat{\mathcal{P}}_{\rm sub}$ via the adiabatic ramp~$\hat{U}$, and overlap of~$\hat{U}\hat{\mathcal{P}}_{\rm sub}\hat{U}^\dagger$ with the ground-state density matrix~$\hat{\rho}$ of the target system.

\subsubsection*{Exact diagonalization}
From the numerical perspective, the ramp-dependence of~$\mathcal{F}_{\rm sub}$ limits our analysis to small system sizes, as the full real-time dynamics of the round-trip protocol must be simulated explicitly.
Here, we analyze an extended toric code with $4 \times 2$ plaquettes and periodic boundaries, see Fig.~\ref{figure-4}a. At time~$t=0$, we initialize the system in~$\ket{\psi_{\rm init}} = \ket{\downarrow}^{\otimes N_{\rm spins}}$, such that $\ket{\psi_{\rm init}}$ is the approximate ground state in the presence of a large external field~$B(t)\sum_j \hat{\sigma}^z_j$ with~$B(0)=5J$; here~$J$ has units of energy and is the strength of the plaquette and vertex term. To obtain a large ground-state fraction, we exponentially ramp down the field~$B(t) \rightarrow 0$ within~$t_{\rm ramp} = 50/J$. Further, we add Wilson loop operators  to the Hamiltonian, i.e., products of Pauli operators around the non-trivial loops of the torus. This ensures we prepare the system in a well-defined topological sector. After applying the instantaneous phase imprinting~$\hat{W}(\phi)$, we perform a backward ramp and measure return fidelities to~$\ket{\psi_{\rm init}}$ (i) globally on all spins $\mathcal{F}_{\rm all}(\phi)$, (ii) on four spins around a plaquette $\mathcal{F}_{\rm plaq}(\phi)$, and (iii) locally on a single spin $\mathcal{F}_{\rm loc}(\phi)$, see Fig.~\ref{figure-4}a. The corresponding coherence distributions are shown and discussed in main text Fig.~\ref{figure-4}b.


\begin{thebibliography}{34}%
\makeatletter
\providecommand \@ifxundefined [1]{%
 \@ifx{#1\undefined}
}%
\providecommand \@ifnum [1]{%
 \ifnum #1\expandafter \@firstoftwo
 \else \expandafter \@secondoftwo
 \fi
}%
\providecommand \@ifx [1]{%
 \ifx #1\expandafter \@firstoftwo
 \else \expandafter \@secondoftwo
 \fi
}%
\providecommand \natexlab [1]{#1}%
\providecommand \enquote  [1]{``#1''}%
\providecommand \bibnamefont  [1]{#1}%
\providecommand \bibfnamefont [1]{#1}%
\providecommand \citenamefont [1]{#1}%
\providecommand \href@noop [0]{\@secondoftwo}%
\providecommand \href [0]{\begingroup \@sanitize@url \@href}%
\providecommand \@href[1]{\@@startlink{#1}\@@href}%
\providecommand \@@href[1]{\endgroup#1\@@endlink}%
\providecommand \@sanitize@url [0]{\catcode `\\12\catcode `\$12\catcode
  `\&12\catcode `\#12\catcode `\^12\catcode `\_12\catcode `\%12\relax}%
\providecommand \@@startlink[1]{}%
\providecommand \@@endlink[0]{}%
\providecommand \url  [0]{\begingroup\@sanitize@url \@url }%
\providecommand \@url [1]{\endgroup\@href {#1}{\urlprefix }}%
\providecommand \urlprefix  [0]{URL }%
\providecommand \Eprint [0]{\href }%
\providecommand \doibase [0]{http://dx.doi.org/}%
\providecommand \selectlanguage [0]{\@gobble}%
\providecommand \bibinfo  [0]{\@secondoftwo}%
\providecommand \bibfield  [0]{\@secondoftwo}%
\providecommand \translation [1]{[#1]}%
\providecommand \BibitemOpen [0]{}%
\providecommand \bibitemStop [0]{}%
\providecommand \bibitemNoStop [0]{.\EOS\space}%
\providecommand \EOS [0]{\spacefactor3000\relax}%
\providecommand \BibitemShut  [1]{\csname bibitem#1\endcsname}%
\let\auto@bib@innerbib\@empty
\bibitem [{\citenamefont {Savary}\ and\ \citenamefont
  {Balents}(2016)}]{Savary2016}%
  \BibitemOpen
  \bibfield  {author} {\bibinfo {author} {\bibfnamefont {Lucile}\ \bibnamefont
  {Savary}}\ and\ \bibinfo {author} {\bibfnamefont {Leon}\ \bibnamefont
  {Balents}},\ }\bibfield  {title} {\enquote {\bibinfo {title} {Quantum spin
  liquids: a review},}\ }\href {https://doi.org/10.1088/0034-4885/80/1/016502}
  {\bibfield  {journal} {\bibinfo  {journal} {Reports on Progress in Physics}\
  }\textbf {\bibinfo {volume} {80}},\ \bibinfo {pages} {016502} (\bibinfo
  {year} {2016})}\BibitemShut {NoStop}%
\bibitem [{\citenamefont {Broholm}\ \emph {et~al.}(2020)\citenamefont
  {Broholm}, \citenamefont {Cava}, \citenamefont {Kivelson}, \citenamefont
  {Nocera}, \citenamefont {Norman},\ and\ \citenamefont
  {Senthil}}]{Broholm2020}%
  \BibitemOpen
  \bibfield  {author} {\bibinfo {author} {\bibfnamefont {C.}~\bibnamefont
  {Broholm}}, \bibinfo {author} {\bibfnamefont {R.~J.}\ \bibnamefont {Cava}},
  \bibinfo {author} {\bibfnamefont {S.~A.}\ \bibnamefont {Kivelson}}, \bibinfo
  {author} {\bibfnamefont {D.~G.}\ \bibnamefont {Nocera}}, \bibinfo {author}
  {\bibfnamefont {M.~R.}\ \bibnamefont {Norman}}, \ and\ \bibinfo {author}
  {\bibfnamefont {T.}~\bibnamefont {Senthil}},\ }\bibfield  {title} {\enquote
  {\bibinfo {title} {Quantum spin liquids},}\ }\href
  {https://doi.org/10.1126/science.aay0668} {\bibfield  {journal} {\bibinfo
  {journal} {Science}\ }\textbf {\bibinfo {volume} {367}} (\bibinfo {year}
  {2020})}\BibitemShut {NoStop}%
\bibitem [{\citenamefont {Wen}(2007)}]{Wen2007}%
  \BibitemOpen
  \bibfield  {author} {\bibinfo {author} {\bibfnamefont {Xiao-Gang}\
  \bibnamefont {Wen}},\ }\href
  {https://doi.org/10.1093/acprof:oso/9780199227259.001.0001} {\emph {\bibinfo
  {title} {Quantum {F}ield {T}heory of {M}any-{B}ody {S}ystems}}}\ (\bibinfo
  {publisher} {Oxford University Press},\ \bibinfo {year} {2007})\BibitemShut
  {NoStop}%
\bibitem [{\citenamefont {Chen}\ \emph {et~al.}(2010)\citenamefont {Chen},
  \citenamefont {Gu},\ and\ \citenamefont {Wen}}]{Chen2010}%
  \BibitemOpen
  \bibfield  {author} {\bibinfo {author} {\bibfnamefont {Xie}\ \bibnamefont
  {Chen}}, \bibinfo {author} {\bibfnamefont {Zheng-Cheng}\ \bibnamefont {Gu}},
  \ and\ \bibinfo {author} {\bibfnamefont {Xiao-Gang}\ \bibnamefont {Wen}},\
  }\bibfield  {title} {\enquote {\bibinfo {title} {{L}ocal unitary
  transformation, long-range quantum entanglement, wave function
  renormalization, and topological order},}\ }\href
  {https://doi.org/10.1103/physrevb.82.155138} {\bibfield  {journal} {\bibinfo
  {journal} {Physical Review B}\ }\textbf {\bibinfo {volume} {82}} (\bibinfo
  {year} {2010})}\BibitemShut {NoStop}%
\bibitem [{\citenamefont {Kitaev}(2003)}]{Kitaev2003}%
  \BibitemOpen
  \bibfield  {author} {\bibinfo {author} {\bibfnamefont {A.~Yu.}\ \bibnamefont
  {Kitaev}},\ }\bibfield  {title} {\enquote {\bibinfo {title} {Fault-tolerant
  quantum computation by anyons},}\ }\href
  {https://doi.org/10.1016/S0003-4916(02)00018-0} {\bibfield  {journal}
  {\bibinfo  {journal} {Ann. Phys. New York}\ }\textbf {\bibinfo {volume}
  {303}},\ \bibinfo {pages} {2--30} (\bibinfo {year} {2003})}\BibitemShut
  {NoStop}%
\bibitem [{\citenamefont {Nayak}\ \emph {et~al.}(2008)\citenamefont {Nayak},
  \citenamefont {Simon}, \citenamefont {Stern}, \citenamefont {Freedman},\ and\
  \citenamefont {Das~Sarma}}]{Nayak2008}%
  \BibitemOpen
  \bibfield  {author} {\bibinfo {author} {\bibfnamefont {Chetan}\ \bibnamefont
  {Nayak}}, \bibinfo {author} {\bibfnamefont {Steven~H.}\ \bibnamefont
  {Simon}}, \bibinfo {author} {\bibfnamefont {Ady}\ \bibnamefont {Stern}},
  \bibinfo {author} {\bibfnamefont {Michael}\ \bibnamefont {Freedman}}, \ and\
  \bibinfo {author} {\bibfnamefont {Sankar}\ \bibnamefont {Das~Sarma}},\
  }\bibfield  {title} {\enquote {\bibinfo {title} {{Non-Abelian anyons and
  topological quantum computation}},}\ }\href
  {https://doi.org/10.1103/revmodphys.80.1083} {\bibfield  {journal} {\bibinfo
  {journal} {Reviews of Modern Physics}\ }\textbf {\bibinfo {volume} {80}},\
  \bibinfo {pages} {1083--1159} (\bibinfo {year} {2008})}\BibitemShut {NoStop}%
\bibitem [{\citenamefont {Semeghini}\ \emph {et~al.}(2021)\citenamefont
  {Semeghini}, \citenamefont {Levine}, \citenamefont {Keesling}, \citenamefont
  {Ebadi}, \citenamefont {Wang}, \citenamefont {Bluvstein}, \citenamefont
  {Verresen}, \citenamefont {Pichler}, \citenamefont {Kalinowski},
  \citenamefont {Samajdar}, \citenamefont {Omran}, \citenamefont {Sachdev},
  \citenamefont {Vishwanath}, \citenamefont {Greiner}, \citenamefont
  {Vuleti{\'{c}}},\ and\ \citenamefont {Lukin}}]{Semeghini2021}%
  \BibitemOpen
  \bibfield  {author} {\bibinfo {author} {\bibfnamefont {G.}~\bibnamefont
  {Semeghini}}, \bibinfo {author} {\bibfnamefont {H.}~\bibnamefont {Levine}},
  \bibinfo {author} {\bibfnamefont {A.}~\bibnamefont {Keesling}}, \bibinfo
  {author} {\bibfnamefont {S.}~\bibnamefont {Ebadi}}, \bibinfo {author}
  {\bibfnamefont {T.~T.}\ \bibnamefont {Wang}}, \bibinfo {author}
  {\bibfnamefont {D.}~\bibnamefont {Bluvstein}}, \bibinfo {author}
  {\bibfnamefont {R.}~\bibnamefont {Verresen}}, \bibinfo {author}
  {\bibfnamefont {H.}~\bibnamefont {Pichler}}, \bibinfo {author} {\bibfnamefont
  {M.}~\bibnamefont {Kalinowski}}, \bibinfo {author} {\bibfnamefont
  {R.}~\bibnamefont {Samajdar}}, \bibinfo {author} {\bibfnamefont
  {A.}~\bibnamefont {Omran}}, \bibinfo {author} {\bibfnamefont
  {S.}~\bibnamefont {Sachdev}}, \bibinfo {author} {\bibfnamefont
  {A.}~\bibnamefont {Vishwanath}}, \bibinfo {author} {\bibfnamefont
  {M.}~\bibnamefont {Greiner}}, \bibinfo {author} {\bibfnamefont
  {V.}~\bibnamefont {Vuleti{\'{c}}}}, \ and\ \bibinfo {author} {\bibfnamefont
  {M.~D.}\ \bibnamefont {Lukin}},\ }\bibfield  {title} {\enquote {\bibinfo
  {title} {Probing topological spin liquids on a programmable quantum
  simulator},}\ }\href {https://doi.org/10.1126/science.abi8794} {\bibfield
  {journal} {\bibinfo  {journal} {Science}\ }\textbf {\bibinfo {volume}
  {374}},\ \bibinfo {pages} {1242--1247} (\bibinfo {year} {2021})}\BibitemShut
  {NoStop}%
\bibitem [{\citenamefont {Satzinger}\ \emph {et~al.}(2021)\citenamefont
  {Satzinger}, \citenamefont {Liu}, \citenamefont {Smith}, \citenamefont
  {Knapp}, \citenamefont {Newman}, \citenamefont {Jones}, \citenamefont {Chen},
  \citenamefont {Quintana}, \citenamefont {Mi}, \citenamefont {Dunsworth},
  \citenamefont {Gidney}, \citenamefont {Aleiner}, \citenamefont {Arute},
  \citenamefont {Arya}, \citenamefont {Atalaya}, \citenamefont {Babbush},
  \citenamefont {Bardin}, \citenamefont {Barends}, \citenamefont {Basso},
  \citenamefont {Bengtsson}, \citenamefont {Bilmes}, \citenamefont {Broughton},
  \citenamefont {Buckley}, \citenamefont {Buell}, \citenamefont {Burkett},
  \citenamefont {Bushnell}, \citenamefont {Chiaro}, \citenamefont {Collins},
  \citenamefont {Courtney}, \citenamefont {Demura}, \citenamefont {Derk},
  \citenamefont {Eppens}, \citenamefont {Erickson}, \citenamefont {Faoro},
  \citenamefont {Farhi}, \citenamefont {Fowler}, \citenamefont {Foxen},
  \citenamefont {Giustina}, \citenamefont {Greene}, \citenamefont {Gross},
  \citenamefont {Harrigan}, \citenamefont {Harrington}, \citenamefont {Hilton},
  \citenamefont {Hong}, \citenamefont {Huang}, \citenamefont {Huggins},
  \citenamefont {Ioffe}, \citenamefont {Isakov}, \citenamefont {Jeffrey},
  \citenamefont {Jiang}, \citenamefont {Kafri}, \citenamefont {Kechedzhi},
  \citenamefont {Khattar}, \citenamefont {Kim}, \citenamefont {Klimov},
  \citenamefont {Korotkov}, \citenamefont {Kostritsa}, \citenamefont
  {Landhuis}, \citenamefont {Laptev}, \citenamefont {Locharla}, \citenamefont
  {Lucero}, \citenamefont {Martin}, \citenamefont {McClean}, \citenamefont
  {McEwen}, \citenamefont {Miao}, \citenamefont {Mohseni}, \citenamefont
  {Montazeri}, \citenamefont {Mruczkiewicz}, \citenamefont {Mutus},
  \citenamefont {Naaman}, \citenamefont {Neeley}, \citenamefont {Neill},
  \citenamefont {Niu}, \citenamefont {O'Brien}, \citenamefont {Opremcak},
  \citenamefont {Pat{\'{o}}}, \citenamefont {Petukhov}, \citenamefont {Rubin},
  \citenamefont {Sank}, \citenamefont {Shvarts}, \citenamefont {Strain},
  \citenamefont {Szalay}, \citenamefont {Villalonga}, \citenamefont {White},
  \citenamefont {Yao}, \citenamefont {Yeh}, \citenamefont {Yoo}, \citenamefont
  {Zalcman}, \citenamefont {Neven}, \citenamefont {Boixo}, \citenamefont
  {Megrant}, \citenamefont {Chen}, \citenamefont {Kelly}, \citenamefont
  {Smelyanskiy}, \citenamefont {Kitaev}, \citenamefont {Knap}, \citenamefont
  {Pollmann},\ and\ \citenamefont {Roushan}}]{Satzinger2021}%
  \BibitemOpen
  \bibfield  {author} {\bibinfo {author} {\bibfnamefont {K.~J.}\ \bibnamefont
  {Satzinger}}, \bibinfo {author} {\bibfnamefont {Y.-J}\ \bibnamefont {Liu}},
  \bibinfo {author} {\bibfnamefont {A.}~\bibnamefont {Smith}}, \bibinfo
  {author} {\bibfnamefont {C.}~\bibnamefont {Knapp}}, \bibinfo {author}
  {\bibfnamefont {M.}~\bibnamefont {Newman}}, \bibinfo {author} {\bibfnamefont
  {C.}~\bibnamefont {Jones}}, \bibinfo {author} {\bibfnamefont
  {Z.}~\bibnamefont {Chen}}, \bibinfo {author} {\bibfnamefont {C.}~\bibnamefont
  {Quintana}}, \bibinfo {author} {\bibfnamefont {X.}~\bibnamefont {Mi}},
  \bibinfo {author} {\bibfnamefont {A.}~\bibnamefont {Dunsworth}}, \bibinfo
  {author} {\bibfnamefont {C.}~\bibnamefont {Gidney}}, \bibinfo {author}
  {\bibfnamefont {I.}~\bibnamefont {Aleiner}}, \bibinfo {author} {\bibfnamefont
  {F.}~\bibnamefont {Arute}}, \bibinfo {author} {\bibfnamefont
  {K.}~\bibnamefont {Arya}}, \bibinfo {author} {\bibfnamefont {J.}~\bibnamefont
  {Atalaya}}, \bibinfo {author} {\bibfnamefont {R.}~\bibnamefont {Babbush}},
  \bibinfo {author} {\bibfnamefont {J.~C.}\ \bibnamefont {Bardin}}, \bibinfo
  {author} {\bibfnamefont {R.}~\bibnamefont {Barends}}, \bibinfo {author}
  {\bibfnamefont {J.}~\bibnamefont {Basso}}, \bibinfo {author} {\bibfnamefont
  {A.}~\bibnamefont {Bengtsson}}, \bibinfo {author} {\bibfnamefont
  {A.}~\bibnamefont {Bilmes}}, \bibinfo {author} {\bibfnamefont
  {M.}~\bibnamefont {Broughton}}, \bibinfo {author} {\bibfnamefont {B.~B.}\
  \bibnamefont {Buckley}}, \bibinfo {author} {\bibfnamefont {D.~A.}\
  \bibnamefont {Buell}}, \bibinfo {author} {\bibfnamefont {B.}~\bibnamefont
  {Burkett}}, \bibinfo {author} {\bibfnamefont {N.}~\bibnamefont {Bushnell}},
  \bibinfo {author} {\bibfnamefont {B.}~\bibnamefont {Chiaro}}, \bibinfo
  {author} {\bibfnamefont {R.}~\bibnamefont {Collins}}, \bibinfo {author}
  {\bibfnamefont {W.}~\bibnamefont {Courtney}}, \bibinfo {author}
  {\bibfnamefont {S.}~\bibnamefont {Demura}}, \bibinfo {author} {\bibfnamefont
  {A.~R.}\ \bibnamefont {Derk}}, \bibinfo {author} {\bibfnamefont
  {D.}~\bibnamefont {Eppens}}, \bibinfo {author} {\bibfnamefont
  {C.}~\bibnamefont {Erickson}}, \bibinfo {author} {\bibfnamefont
  {L.}~\bibnamefont {Faoro}}, \bibinfo {author} {\bibfnamefont
  {E.}~\bibnamefont {Farhi}}, \bibinfo {author} {\bibfnamefont {A.~G.}\
  \bibnamefont {Fowler}}, \bibinfo {author} {\bibfnamefont {B.}~\bibnamefont
  {Foxen}}, \bibinfo {author} {\bibfnamefont {M.}~\bibnamefont {Giustina}},
  \bibinfo {author} {\bibfnamefont {A.}~\bibnamefont {Greene}}, \bibinfo
  {author} {\bibfnamefont {J.~A.}\ \bibnamefont {Gross}}, \bibinfo {author}
  {\bibfnamefont {M.~P.}\ \bibnamefont {Harrigan}}, \bibinfo {author}
  {\bibfnamefont {S.~D.}\ \bibnamefont {Harrington}}, \bibinfo {author}
  {\bibfnamefont {J.}~\bibnamefont {Hilton}}, \bibinfo {author} {\bibfnamefont
  {S.}~\bibnamefont {Hong}}, \bibinfo {author} {\bibfnamefont {T.}~\bibnamefont
  {Huang}}, \bibinfo {author} {\bibfnamefont {W.~J.}\ \bibnamefont {Huggins}},
  \bibinfo {author} {\bibfnamefont {L.~B.}\ \bibnamefont {Ioffe}}, \bibinfo
  {author} {\bibfnamefont {S.~V.}\ \bibnamefont {Isakov}}, \bibinfo {author}
  {\bibfnamefont {E.}~\bibnamefont {Jeffrey}}, \bibinfo {author} {\bibfnamefont
  {Z.}~\bibnamefont {Jiang}}, \bibinfo {author} {\bibfnamefont
  {D.}~\bibnamefont {Kafri}}, \bibinfo {author} {\bibfnamefont
  {K.}~\bibnamefont {Kechedzhi}}, \bibinfo {author} {\bibfnamefont
  {T.}~\bibnamefont {Khattar}}, \bibinfo {author} {\bibfnamefont
  {S.}~\bibnamefont {Kim}}, \bibinfo {author} {\bibfnamefont {P.~V.}\
  \bibnamefont {Klimov}}, \bibinfo {author} {\bibfnamefont {A.~N.}\
  \bibnamefont {Korotkov}}, \bibinfo {author} {\bibfnamefont {F.}~\bibnamefont
  {Kostritsa}}, \bibinfo {author} {\bibfnamefont {D.}~\bibnamefont {Landhuis}},
  \bibinfo {author} {\bibfnamefont {P.}~\bibnamefont {Laptev}}, \bibinfo
  {author} {\bibfnamefont {A.}~\bibnamefont {Locharla}}, \bibinfo {author}
  {\bibfnamefont {E.}~\bibnamefont {Lucero}}, \bibinfo {author} {\bibfnamefont
  {O.}~\bibnamefont {Martin}}, \bibinfo {author} {\bibfnamefont {J.~R.}\
  \bibnamefont {McClean}}, \bibinfo {author} {\bibfnamefont {M.}~\bibnamefont
  {McEwen}}, \bibinfo {author} {\bibfnamefont {K.~C.}\ \bibnamefont {Miao}},
  \bibinfo {author} {\bibfnamefont {M.}~\bibnamefont {Mohseni}}, \bibinfo
  {author} {\bibfnamefont {S.}~\bibnamefont {Montazeri}}, \bibinfo {author}
  {\bibfnamefont {W.}~\bibnamefont {Mruczkiewicz}}, \bibinfo {author}
  {\bibfnamefont {J.}~\bibnamefont {Mutus}}, \bibinfo {author} {\bibfnamefont
  {O.}~\bibnamefont {Naaman}}, \bibinfo {author} {\bibfnamefont
  {M.}~\bibnamefont {Neeley}}, \bibinfo {author} {\bibfnamefont
  {C.}~\bibnamefont {Neill}}, \bibinfo {author} {\bibfnamefont {M.~Y.}\
  \bibnamefont {Niu}}, \bibinfo {author} {\bibfnamefont {T.~E.}\ \bibnamefont
  {O'Brien}}, \bibinfo {author} {\bibfnamefont {A.}~\bibnamefont {Opremcak}},
  \bibinfo {author} {\bibfnamefont {B.}~\bibnamefont {Pat{\'{o}}}}, \bibinfo
  {author} {\bibfnamefont {A.}~\bibnamefont {Petukhov}}, \bibinfo {author}
  {\bibfnamefont {N.~C.}\ \bibnamefont {Rubin}}, \bibinfo {author}
  {\bibfnamefont {D.}~\bibnamefont {Sank}}, \bibinfo {author} {\bibfnamefont
  {V.}~\bibnamefont {Shvarts}}, \bibinfo {author} {\bibfnamefont
  {D.}~\bibnamefont {Strain}}, \bibinfo {author} {\bibfnamefont
  {M.}~\bibnamefont {Szalay}}, \bibinfo {author} {\bibfnamefont
  {B.}~\bibnamefont {Villalonga}}, \bibinfo {author} {\bibfnamefont {T.~C.}\
  \bibnamefont {White}}, \bibinfo {author} {\bibfnamefont {Z.}~\bibnamefont
  {Yao}}, \bibinfo {author} {\bibfnamefont {P.}~\bibnamefont {Yeh}}, \bibinfo
  {author} {\bibfnamefont {J.}~\bibnamefont {Yoo}}, \bibinfo {author}
  {\bibfnamefont {A.}~\bibnamefont {Zalcman}}, \bibinfo {author} {\bibfnamefont
  {H.}~\bibnamefont {Neven}}, \bibinfo {author} {\bibfnamefont
  {S.}~\bibnamefont {Boixo}}, \bibinfo {author} {\bibfnamefont
  {A.}~\bibnamefont {Megrant}}, \bibinfo {author} {\bibfnamefont
  {Y.}~\bibnamefont {Chen}}, \bibinfo {author} {\bibfnamefont {J.}~\bibnamefont
  {Kelly}}, \bibinfo {author} {\bibfnamefont {V.}~\bibnamefont {Smelyanskiy}},
  \bibinfo {author} {\bibfnamefont {A.}~\bibnamefont {Kitaev}}, \bibinfo
  {author} {\bibfnamefont {M.}~\bibnamefont {Knap}}, \bibinfo {author}
  {\bibfnamefont {F.}~\bibnamefont {Pollmann}}, \ and\ \bibinfo {author}
  {\bibfnamefont {P.}~\bibnamefont {Roushan}},\ }\bibfield  {title} {\enquote
  {\bibinfo {title} {Realizing topologically ordered states on a quantum
  processor},}\ }\href {\doibase 10.1126/science.abi8378} {\bibfield  {journal}
  {\bibinfo  {journal} {Science}\ }\textbf {\bibinfo {volume} {374}},\ \bibinfo
  {pages} {1237--1241} (\bibinfo {year} {2021})}\BibitemShut {NoStop}%
\bibitem [{\citenamefont {Bornet}\ \emph {et~al.}(2026)\citenamefont {Bornet},
  \citenamefont {Bintz}, \citenamefont {Chen}, \citenamefont {Emperauger},
  \citenamefont {Qiao}, \citenamefont {Martin}, \citenamefont {Barredo},
  \citenamefont {Chatterjee}, \citenamefont {Liu}, \citenamefont {Lahaye},
  \citenamefont {Zaletel}, \citenamefont {Yao},\ and\ \citenamefont
  {Browaeys}}]{Bornet2026}%
  \BibitemOpen
  \bibfield  {author} {\bibinfo {author} {\bibfnamefont {Guillaume}\
  \bibnamefont {Bornet}}, \bibinfo {author} {\bibfnamefont {Marcus}\
  \bibnamefont {Bintz}}, \bibinfo {author} {\bibfnamefont {Cheng}\ \bibnamefont
  {Chen}}, \bibinfo {author} {\bibfnamefont {Gabriel}\ \bibnamefont
  {Emperauger}}, \bibinfo {author} {\bibfnamefont {Mu}~\bibnamefont {Qiao}},
  \bibinfo {author} {\bibfnamefont {Romain}\ \bibnamefont {Martin}}, \bibinfo
  {author} {\bibfnamefont {Daniel}\ \bibnamefont {Barredo}}, \bibinfo {author}
  {\bibfnamefont {Shubhayu}\ \bibnamefont {Chatterjee}}, \bibinfo {author}
  {\bibfnamefont {Vincent~S.}\ \bibnamefont {Liu}}, \bibinfo {author}
  {\bibfnamefont {Thierry}\ \bibnamefont {Lahaye}}, \bibinfo {author}
  {\bibfnamefont {Michael~P.}\ \bibnamefont {Zaletel}}, \bibinfo {author}
  {\bibfnamefont {Norman~Y.}\ \bibnamefont {Yao}}, \ and\ \bibinfo {author}
  {\bibfnamefont {Antoine}\ \bibnamefont {Browaeys}},\ }\href
  {https://arxiv.org/pdf/2602.14323.pdf} {\enquote {\bibinfo {title} {{Dirac
  Spin Liquid Candidate in a Rydberg Quantum Simulator}},}\ } (\bibinfo {year}
  {2026}),\ \Eprint {http://arxiv.org/abs/2602.14323} {arXiv:2602.14323}
  \BibitemShut {NoStop}%
\bibitem [{\citenamefont {Geim}\ \emph {et~al.}(2026)\citenamefont {Geim},
  \citenamefont {Koyluoglu}, \citenamefont {Evered}, \citenamefont {Sahay},
  \citenamefont {Li}, \citenamefont {Xu}, \citenamefont {Bluvstein},
  \citenamefont {Gjonbalaj}, \citenamefont {Maskara}, \citenamefont
  {Kalinowski}, \citenamefont {Manovitz}, \citenamefont {Verresen},
  \citenamefont {Yelin}, \citenamefont {Feldmeier}, \citenamefont {Greiner},
  \citenamefont {Vuletic},\ and\ \citenamefont {Lukin}}]{Geim2026}%
  \BibitemOpen
  \bibfield  {author} {\bibinfo {author} {\bibfnamefont {Alexandra~A.}\
  \bibnamefont {Geim}}, \bibinfo {author} {\bibfnamefont {Nazli~Ugur}\
  \bibnamefont {Koyluoglu}}, \bibinfo {author} {\bibfnamefont {Simon~J.}\
  \bibnamefont {Evered}}, \bibinfo {author} {\bibfnamefont {Rahul}\
  \bibnamefont {Sahay}}, \bibinfo {author} {\bibfnamefont {Sophie~H.}\
  \bibnamefont {Li}}, \bibinfo {author} {\bibfnamefont {Muqing}\ \bibnamefont
  {Xu}}, \bibinfo {author} {\bibfnamefont {Dolev}\ \bibnamefont {Bluvstein}},
  \bibinfo {author} {\bibfnamefont {Nik~O.}\ \bibnamefont {Gjonbalaj}},
  \bibinfo {author} {\bibfnamefont {Nishad}\ \bibnamefont {Maskara}}, \bibinfo
  {author} {\bibfnamefont {Marcin}\ \bibnamefont {Kalinowski}}, \bibinfo
  {author} {\bibfnamefont {Tom}\ \bibnamefont {Manovitz}}, \bibinfo {author}
  {\bibfnamefont {Ruben}\ \bibnamefont {Verresen}}, \bibinfo {author}
  {\bibfnamefont {Susanne~F.}\ \bibnamefont {Yelin}}, \bibinfo {author}
  {\bibfnamefont {Johannes}\ \bibnamefont {Feldmeier}}, \bibinfo {author}
  {\bibfnamefont {Markus}\ \bibnamefont {Greiner}}, \bibinfo {author}
  {\bibfnamefont {Vladan}\ \bibnamefont {Vuletic}}, \ and\ \bibinfo {author}
  {\bibfnamefont {Mikhail~D.}\ \bibnamefont {Lukin}},\ }\href
  {https://arxiv.org/pdf/2602.18555.pdf} {\enquote {\bibinfo {title}
  {{Engineering quantum criticality and dynamics on an analog-digital
  simulator}},}\ } (\bibinfo {year} {2026}),\ \Eprint
  {http://arxiv.org/abs/2602.18555} {arXiv:2602.18555} \BibitemShut {NoStop}%
\bibitem [{\citenamefont {Karch}\ \emph {et~al.}(2026)\citenamefont {Karch},
  \citenamefont {Will}, \citenamefont {Rodriguez}, \citenamefont {Liebster},
  \citenamefont {Huh}, \citenamefont {Knap}, \citenamefont {Pollmann},
  \citenamefont {Kuhlenkamp}, \citenamefont {Bloch},\ and\ \citenamefont
  {Aidelsburger}}]{Karch2026}%
  \BibitemOpen
  \bibfield  {author} {\bibinfo {author} {\bibfnamefont {Simon}\ \bibnamefont
  {Karch}}, \bibinfo {author} {\bibfnamefont {Melissa}\ \bibnamefont {Will}},
  \bibinfo {author} {\bibfnamefont {Irene~Prieto}\ \bibnamefont {Rodriguez}},
  \bibinfo {author} {\bibfnamefont {Nikolas}\ \bibnamefont {Liebster}},
  \bibinfo {author} {\bibfnamefont {SeungJung}\ \bibnamefont {Huh}}, \bibinfo
  {author} {\bibfnamefont {Michael}\ \bibnamefont {Knap}}, \bibinfo {author}
  {\bibfnamefont {Frank}\ \bibnamefont {Pollmann}}, \bibinfo {author}
  {\bibfnamefont {Clemens}\ \bibnamefont {Kuhlenkamp}}, \bibinfo {author}
  {\bibfnamefont {Immanuel}\ \bibnamefont {Bloch}}, \ and\ \bibinfo {author}
  {\bibfnamefont {Monika}\ \bibnamefont {Aidelsburger}},\ }\href
  {https://arxiv.org/pdf/2604.24744.pdf} {\enquote {\bibinfo {title}
  {{Dynamical preparation of U(1) quantum spin liquids in an analogue quantum
  simulator}},}\ } (\bibinfo {year} {2026}),\ \Eprint
  {http://arxiv.org/abs/2604.24744} {arXiv:2604.24744} \BibitemShut {NoStop}%
\bibitem [{\citenamefont {Gregor}\ \emph {et~al.}(2011)\citenamefont {Gregor},
  \citenamefont {Huse}, \citenamefont {Moessner},\ and\ \citenamefont
  {Sondhi}}]{Gregor2011}%
  \BibitemOpen
  \bibfield  {author} {\bibinfo {author} {\bibfnamefont {K.}~\bibnamefont
  {Gregor}}, \bibinfo {author} {\bibfnamefont {David~A.}\ \bibnamefont {Huse}},
  \bibinfo {author} {\bibfnamefont {R.}~\bibnamefont {Moessner}}, \ and\
  \bibinfo {author} {\bibfnamefont {S.~L.}\ \bibnamefont {Sondhi}},\ }\bibfield
   {title} {\enquote {\bibinfo {title} {Diagnosing deconfinement and
  topological order},}\ }\href {https://doi.org/10.1088/1367-2630/13/2/025009}
  {\bibfield  {journal} {\bibinfo  {journal} {New Journal of Physics}\ }\textbf
  {\bibinfo {volume} {13}},\ \bibinfo {pages} {025009} (\bibinfo {year}
  {2011})}\BibitemShut {NoStop}%
\bibitem [{\citenamefont {Verresen}\ \emph {et~al.}(2021)\citenamefont
  {Verresen}, \citenamefont {Lukin},\ and\ \citenamefont
  {Vishwanath}}]{Verresen2021}%
  \BibitemOpen
  \bibfield  {author} {\bibinfo {author} {\bibfnamefont {Ruben}\ \bibnamefont
  {Verresen}}, \bibinfo {author} {\bibfnamefont {Mikhail~D.}\ \bibnamefont
  {Lukin}}, \ and\ \bibinfo {author} {\bibfnamefont {Ashvin}\ \bibnamefont
  {Vishwanath}},\ }\bibfield  {title} {\enquote {\bibinfo {title} {Prediction
  of {T}oric {C}ode {T}opological {O}rder from {R}ydberg {B}lockade},}\ }\href
  {https://doi.org/10.1103/physrevx.11.031005} {\bibfield  {journal} {\bibinfo
  {journal} {Physical Review X}\ }\textbf {\bibinfo {volume} {11}},\ \bibinfo
  {pages} {031005} (\bibinfo {year} {2021})}\BibitemShut {NoStop}%
\bibitem [{\citenamefont {Wang}\ and\ \citenamefont {Pollet}(2025)}]{Wang2025}%
  \BibitemOpen
  \bibfield  {author} {\bibinfo {author} {\bibfnamefont {Zhenjiu}\ \bibnamefont
  {Wang}}\ and\ \bibinfo {author} {\bibfnamefont {Lode}\ \bibnamefont
  {Pollet}},\ }\bibfield  {title} {\enquote {\bibinfo {title} {{Renormalized
  Classical Spin Liquid on the Ruby Lattice}},}\ }\href
  {https://doi.org/10.1103/physrevlett.134.086601} {\bibfield  {journal}
  {\bibinfo  {journal} {Physical Review Letters}\ }\textbf {\bibinfo {volume}
  {134}},\ \bibinfo {pages} {086601} (\bibinfo {year} {2025})}\BibitemShut
  {NoStop}%
\bibitem [{\citenamefont {Baum}\ \emph {et~al.}(1985)\citenamefont {Baum},
  \citenamefont {Munowitz}, \citenamefont {Garroway},\ and\ \citenamefont
  {Pines}}]{Baum1985}%
  \BibitemOpen
  \bibfield  {author} {\bibinfo {author} {\bibfnamefont {J.}~\bibnamefont
  {Baum}}, \bibinfo {author} {\bibfnamefont {M.}~\bibnamefont {Munowitz}},
  \bibinfo {author} {\bibfnamefont {A.~N.}\ \bibnamefont {Garroway}}, \ and\
  \bibinfo {author} {\bibfnamefont {A.}~\bibnamefont {Pines}},\ }\bibfield
  {title} {\enquote {\bibinfo {title} {{Multiple-quantum dynamics in solid
  state NMR}},}\ }\href {https://doi.org/10.1063/1.449344} {\bibfield
  {journal} {\bibinfo  {journal} {The Journal of Chemical Physics}\ }\textbf
  {\bibinfo {volume} {83}},\ \bibinfo {pages} {2015--2025} (\bibinfo {year}
  {1985})}\BibitemShut {NoStop}%
\bibitem [{\citenamefont {Lewis-Swan}\ \emph {et~al.}(2020)\citenamefont
  {Lewis-Swan}, \citenamefont {Muleady},\ and\ \citenamefont
  {Rey}}]{LewisSwan2020}%
  \BibitemOpen
  \bibfield  {author} {\bibinfo {author} {\bibfnamefont {R.~J.}\ \bibnamefont
  {Lewis-Swan}}, \bibinfo {author} {\bibfnamefont {S.~R.}\ \bibnamefont
  {Muleady}}, \ and\ \bibinfo {author} {\bibfnamefont {A.~M.}\ \bibnamefont
  {Rey}},\ }\bibfield  {title} {\enquote {\bibinfo {title} {{Detecting
  Out-of-Time-Order Correlations via Quasiadiabatic Echoes as a Tool to Reveal
  Quantum Coherence in Equilibrium Quantum Phase Transitions}},}\ }\href
  {https://doi.org/10.1103/physrevlett.125.240605} {\bibfield  {journal}
  {\bibinfo  {journal} {Physical Review Letters}\ }\textbf {\bibinfo {volume}
  {125}},\ \bibinfo {pages} {240605} (\bibinfo {year} {2020})}\BibitemShut
  {NoStop}%
\bibitem [{\citenamefont {Gärttner}\ \emph {et~al.}(2017)\citenamefont
  {Gärttner}, \citenamefont {Bohnet}, \citenamefont {Safavi-Naini},
  \citenamefont {Wall}, \citenamefont {Bollinger},\ and\ \citenamefont
  {Rey}}]{Gaerttner2017}%
  \BibitemOpen
  \bibfield  {author} {\bibinfo {author} {\bibfnamefont {Martin}\ \bibnamefont
  {Gärttner}}, \bibinfo {author} {\bibfnamefont {Justin~G.}\ \bibnamefont
  {Bohnet}}, \bibinfo {author} {\bibfnamefont {Arghavan}\ \bibnamefont
  {Safavi-Naini}}, \bibinfo {author} {\bibfnamefont {Michael~L.}\ \bibnamefont
  {Wall}}, \bibinfo {author} {\bibfnamefont {John~J.}\ \bibnamefont
  {Bollinger}}, \ and\ \bibinfo {author} {\bibfnamefont {Ana~Maria}\
  \bibnamefont {Rey}},\ }\bibfield  {title} {\enquote {\bibinfo {title}
  {Measuring out-of-time-order correlations and multiple quantum spectra in a
  trapped-ion quantum magnet},}\ }\href {https://doi.org/10.1038/nphys4119}
  {\bibfield  {journal} {\bibinfo  {journal} {Nature Physics}\ }\textbf
  {\bibinfo {volume} {13}},\ \bibinfo {pages} {781--786} (\bibinfo {year}
  {2017})}\BibitemShut {NoStop}%
\bibitem [{\citenamefont {Gärttner}\ \emph {et~al.}(2018)\citenamefont
  {Gärttner}, \citenamefont {Hauke},\ and\ \citenamefont
  {Rey}}]{Gaerttner2018}%
  \BibitemOpen
  \bibfield  {author} {\bibinfo {author} {\bibfnamefont {Martin}\ \bibnamefont
  {Gärttner}}, \bibinfo {author} {\bibfnamefont {Philipp}\ \bibnamefont
  {Hauke}}, \ and\ \bibinfo {author} {\bibfnamefont {Ana~Maria}\ \bibnamefont
  {Rey}},\ }\bibfield  {title} {\enquote {\bibinfo {title} {{Relating
  Out-of-Time-Order Correlations to Entanglement via Multiple-Quantum
  Coherences}},}\ }\href {https://doi.org/10.1103/physrevlett.120.040402}
  {\bibfield  {journal} {\bibinfo  {journal} {Physical Review Letters}\
  }\textbf {\bibinfo {volume} {120}},\ \bibinfo {pages} {040402} (\bibinfo
  {year} {2018})}\BibitemShut {NoStop}%
\bibitem [{\citenamefont {Linsel}\ and\ \citenamefont
  {Pollet}(2026{\natexlab{a}})}]{Linsel2026}%
  \BibitemOpen
  \bibfield  {author} {\bibinfo {author} {\bibfnamefont {Simon~M.}\
  \bibnamefont {Linsel}}\ and\ \bibinfo {author} {\bibfnamefont {Lode}\
  \bibnamefont {Pollet}},\ }\bibfield  {title} {\enquote {\bibinfo {title}
  {{ParaToric 1.0: Continuous-time quantum Monte Carlo for the toric code in a
  parallel field}},}\ }\href {https://doi.org/10.21468/scipostphyscodeb.75}
  {\bibfield  {journal} {\bibinfo  {journal} {SciPost Physics Codebases}\ ,\
  \bibinfo {pages} {75}} (\bibinfo {year} {2026}{\natexlab{a}})}\BibitemShut
  {NoStop}%
\bibitem [{\citenamefont {Linsel}\ and\ \citenamefont
  {Pollet}(2026{\natexlab{b}})}]{Linsel2026a}%
  \BibitemOpen
  \bibfield  {author} {\bibinfo {author} {\bibfnamefont {Simon~M.}\
  \bibnamefont {Linsel}}\ and\ \bibinfo {author} {\bibfnamefont {Lode}\
  \bibnamefont {Pollet}},\ }\bibfield  {title} {\enquote {\bibinfo {title}
  {{Codebase release 1.0 for ParaToric}},}\ }\href
  {https://doi.org/10.21468/scipostphyscodeb.75-r1.0} {\bibfield  {journal}
  {\bibinfo  {journal} {SciPost Physics Codebases}\ ,\ \bibinfo {pages}
  {75--r1.0}} (\bibinfo {year} {2026}{\natexlab{b}})}\BibitemShut {NoStop}%
\bibitem [{\citenamefont {Trebst}\ \emph {et~al.}(2007)\citenamefont {Trebst},
  \citenamefont {Werner}, \citenamefont {Troyer}, \citenamefont {Shtengel},\
  and\ \citenamefont {Nayak}}]{Trebst2007}%
  \BibitemOpen
  \bibfield  {author} {\bibinfo {author} {\bibfnamefont {Simon}\ \bibnamefont
  {Trebst}}, \bibinfo {author} {\bibfnamefont {Philipp}\ \bibnamefont
  {Werner}}, \bibinfo {author} {\bibfnamefont {Matthias}\ \bibnamefont
  {Troyer}}, \bibinfo {author} {\bibfnamefont {Kirill}\ \bibnamefont
  {Shtengel}}, \ and\ \bibinfo {author} {\bibfnamefont {Chetan}\ \bibnamefont
  {Nayak}},\ }\bibfield  {title} {\enquote {\bibinfo {title} {Breakdown of a
  {T}opological {P}hase: {Q}uantum {P}hase {T}ransition in a {L}oop {G}as
  {M}odel with {T}ension},}\ }\href
  {https://doi.org/10.1103/physrevlett.98.070602} {\bibfield  {journal}
  {\bibinfo  {journal} {Physical Review Letters}\ }\textbf {\bibinfo {volume}
  {98}} (\bibinfo {year} {2007})}\BibitemShut {NoStop}%
\bibitem [{\citenamefont {Wu}\ \emph {et~al.}(2012)\citenamefont {Wu},
  \citenamefont {Deng},\ and\ \citenamefont {Prokof’ev}}]{Wu2012}%
  \BibitemOpen
  \bibfield  {author} {\bibinfo {author} {\bibfnamefont {Fengcheng}\
  \bibnamefont {Wu}}, \bibinfo {author} {\bibfnamefont {Youjin}\ \bibnamefont
  {Deng}}, \ and\ \bibinfo {author} {\bibfnamefont {Nikolay}\ \bibnamefont
  {Prokof’ev}},\ }\bibfield  {title} {\enquote {\bibinfo {title} {Phase
  diagram of the toric code model in a parallel magnetic field},}\ }\href
  {https://doi.org/10.1103/physrevb.85.195104} {\bibfield  {journal} {\bibinfo
  {journal} {Physical Review B}\ }\textbf {\bibinfo {volume} {85}},\ \bibinfo
  {pages} {195104} (\bibinfo {year} {2012})}\BibitemShut {NoStop}%
\bibitem [{\citenamefont {Wegner}(1971)}]{Wegner1971}%
  \BibitemOpen
  \bibfield  {author} {\bibinfo {author} {\bibfnamefont {Franz~J.}\
  \bibnamefont {Wegner}},\ }\bibfield  {title} {\enquote {\bibinfo {title}
  {Duality in {G}eneralized {I}sing {M}odels and {P}hase {T}ransitions without
  {L}ocal {O}rder {P}arameters},}\ }\href {https://doi.org/10.1063/1.1665530}
  {\bibfield  {journal} {\bibinfo  {journal} {Journal of Mathematical Physics}\
  }\textbf {\bibinfo {volume} {12}},\ \bibinfo {pages} {2259--2272} (\bibinfo
  {year} {1971})}\BibitemShut {NoStop}%
\bibitem [{\citenamefont {Kogut}(1979)}]{Kogut1979}%
  \BibitemOpen
  \bibfield  {author} {\bibinfo {author} {\bibfnamefont {John~B.}\ \bibnamefont
  {Kogut}},\ }\bibfield  {title} {\enquote {\bibinfo {title} {An introduction
  to lattice gauge theory and spin systems},}\ }\href
  {https://doi.org/10.1103/revmodphys.51.659} {\bibfield  {journal} {\bibinfo
  {journal} {Reviews of Modern Physics}\ }\textbf {\bibinfo {volume} {51}},\
  \bibinfo {pages} {659--713} (\bibinfo {year} {1979})}\BibitemShut {NoStop}%
\bibitem [{\citenamefont {Fradkin}\ and\ \citenamefont
  {Shenker}(1979)}]{FradkinShenker1979}%
  \BibitemOpen
  \bibfield  {author} {\bibinfo {author} {\bibfnamefont {Eduardo}\ \bibnamefont
  {Fradkin}}\ and\ \bibinfo {author} {\bibfnamefont {Stephen~H.}\ \bibnamefont
  {Shenker}},\ }\bibfield  {title} {\enquote {\bibinfo {title} {Phase diagrams
  of lattice gauge theories with {H}iggs fields},}\ }\href
  {https://doi.org/10.1103/physrevd.19.3682} {\bibfield  {journal} {\bibinfo
  {journal} {Physical Review D}\ }\textbf {\bibinfo {volume} {19}},\ \bibinfo
  {pages} {3682--3697} (\bibinfo {year} {1979})}\BibitemShut {NoStop}%
\bibitem [{\citenamefont {Fredenhagen}\ and\ \citenamefont
  {Marcu}(1986)}]{Fredenhagen1986}%
  \BibitemOpen
  \bibfield  {author} {\bibinfo {author} {\bibfnamefont {Klaus}\ \bibnamefont
  {Fredenhagen}}\ and\ \bibinfo {author} {\bibfnamefont {Mihail}\ \bibnamefont
  {Marcu}},\ }\bibfield  {title} {\enquote {\bibinfo {title} {{Confinement
  criterion for QCD with dynamical quarks}},}\ }\href
  {https://doi.org/10.1103/physrevlett.56.223} {\bibfield  {journal} {\bibinfo
  {journal} {Physical Review Letters}\ }\textbf {\bibinfo {volume} {56}},\
  \bibinfo {pages} {223--224} (\bibinfo {year} {1986})}\BibitemShut {NoStop}%
\bibitem [{\citenamefont {Linsel}\ \emph {et~al.}(2024)\citenamefont {Linsel},
  \citenamefont {Bohrdt}, \citenamefont {Homeier}, \citenamefont {Pollet},\
  and\ \citenamefont {Grusdt}}]{Linsel2024}%
  \BibitemOpen
  \bibfield  {author} {\bibinfo {author} {\bibfnamefont {Simon~M.}\
  \bibnamefont {Linsel}}, \bibinfo {author} {\bibfnamefont {Annabelle}\
  \bibnamefont {Bohrdt}}, \bibinfo {author} {\bibfnamefont {Lukas}\
  \bibnamefont {Homeier}}, \bibinfo {author} {\bibfnamefont {Lode}\
  \bibnamefont {Pollet}}, \ and\ \bibinfo {author} {\bibfnamefont {Fabian}\
  \bibnamefont {Grusdt}},\ }\bibfield  {title} {\enquote {\bibinfo {title}
  {{Percolation as a confinement order parameter in Z2 lattice gauge
  theories}},}\ }\href {https://doi.org/10.1103/physrevb.110.l241101}
  {\bibfield  {journal} {\bibinfo  {journal} {Physical Review B}\ }\textbf
  {\bibinfo {volume} {110}},\ \bibinfo {pages} {l241101} (\bibinfo {year}
  {2024})}\BibitemShut {NoStop}%
\bibitem [{\citenamefont {Linsel}\ \emph {et~al.}(2026)\citenamefont {Linsel},
  \citenamefont {Pollet},\ and\ \citenamefont {Grusdt}}]{Linsel2026_prxq}%
  \BibitemOpen
  \bibfield  {author} {\bibinfo {author} {\bibfnamefont {Simon~M.}\
  \bibnamefont {Linsel}}, \bibinfo {author} {\bibfnamefont {Lode}\ \bibnamefont
  {Pollet}}, \ and\ \bibinfo {author} {\bibfnamefont {Fabian}\ \bibnamefont
  {Grusdt}},\ }\bibfield  {title} {\enquote {\bibinfo {title} {{Independent e-
  and m-Anyon Confinement in the Parallel Field Toric Code on Non-Square
  Lattices}},}\ }\href {https://doi.org/10.1103/gtth-cclr} {\bibfield
  {journal} {\bibinfo  {journal} {PRX Quantum}\ }\textbf {\bibinfo {volume}
  {7}} (\bibinfo {year} {2026})}\BibitemShut {NoStop}%
\bibitem [{\citenamefont {Sahay}\ \emph {et~al.}(2022)\citenamefont {Sahay},
  \citenamefont {Vishwanath},\ and\ \citenamefont {Verrese}}]{Sahay2022}%
  \BibitemOpen
  \bibfield  {author} {\bibinfo {author} {\bibfnamefont {Rahul}\ \bibnamefont
  {Sahay}}, \bibinfo {author} {\bibfnamefont {Ashvin}\ \bibnamefont
  {Vishwanath}}, \ and\ \bibinfo {author} {\bibfnamefont {Ruben}\ \bibnamefont
  {Verrese}},\ }\href {https://arxiv.org/pdf/2211.01381.pdf} {\enquote
  {\bibinfo {title} {{Quantum Spin Puddles and Lakes: NISQ-Era Spin Liquids
  from Non-Equilibrium Dynamics}},}\ } (\bibinfo {year} {2022}),\ \Eprint
  {http://arxiv.org/abs/2211.013814} {arXiv:2211.013814} \BibitemShut {NoStop}%
\bibitem [{\citenamefont {Giudici}\ \emph {et~al.}(2022)\citenamefont
  {Giudici}, \citenamefont {Lukin},\ and\ \citenamefont
  {Pichler}}]{Giudici2022}%
  \BibitemOpen
  \bibfield  {author} {\bibinfo {author} {\bibfnamefont {Giuliano}\
  \bibnamefont {Giudici}}, \bibinfo {author} {\bibfnamefont {Mikhail~D.}\
  \bibnamefont {Lukin}}, \ and\ \bibinfo {author} {\bibfnamefont {Hannes}\
  \bibnamefont {Pichler}},\ }\bibfield  {title} {\enquote {\bibinfo {title}
  {Dynamical {P}reparation of {Q}uantum {S}pin {L}iquids in {R}ydberg {A}tom
  {A}rrays},}\ }\href {https://doi.org/10.1103/physrevlett.129.090401}
  {\bibfield  {journal} {\bibinfo  {journal} {Physical Review Letters}\
  }\textbf {\bibinfo {volume} {129}},\ \bibinfo {pages} {090401} (\bibinfo
  {year} {2022})}\BibitemShut {NoStop}%
\bibitem [{\citenamefont {Mauron}\ \emph {et~al.}(2025)\citenamefont {Mauron},
  \citenamefont {Denis}, \citenamefont {Nys},\ and\ \citenamefont
  {Carleo}}]{Mauron2025}%
  \BibitemOpen
  \bibfield  {author} {\bibinfo {author} {\bibfnamefont {Linda}\ \bibnamefont
  {Mauron}}, \bibinfo {author} {\bibfnamefont {Zakari}\ \bibnamefont {Denis}},
  \bibinfo {author} {\bibfnamefont {Jannes}\ \bibnamefont {Nys}}, \ and\
  \bibinfo {author} {\bibfnamefont {Giuseppe}\ \bibnamefont {Carleo}},\
  }\bibfield  {title} {\enquote {\bibinfo {title} {{Predicting topological
  entanglement entropy in a Rydberg analogue simulator}},}\ }\href
  {https://doi.org/10.1038/s41567-025-02944-3} {\bibfield  {journal} {\bibinfo
  {journal} {Nature Physics}\ }\textbf {\bibinfo {volume} {21}},\ \bibinfo
  {pages} {1332--1337} (\bibinfo {year} {2025})}\BibitemShut {NoStop}%
\bibitem [{\citenamefont {Karch}\ \emph {et~al.}(2025)\citenamefont {Karch},
  \citenamefont {Bandyopadhyay}, \citenamefont {Sun}, \citenamefont {Impertro},
  \citenamefont {Huh}, \citenamefont {Rodríguez}, \citenamefont {Wienand},
  \citenamefont {Ketterle}, \citenamefont {Heyl}, \citenamefont {Polkovnikov},
  \citenamefont {Bloch},\ and\ \citenamefont {Aidelsburger}}]{Karch2025}%
  \BibitemOpen
  \bibfield  {author} {\bibinfo {author} {\bibfnamefont {Simon}\ \bibnamefont
  {Karch}}, \bibinfo {author} {\bibfnamefont {Souvik}\ \bibnamefont
  {Bandyopadhyay}}, \bibinfo {author} {\bibfnamefont {Zheng-Hang}\ \bibnamefont
  {Sun}}, \bibinfo {author} {\bibfnamefont {Alexander}\ \bibnamefont
  {Impertro}}, \bibinfo {author} {\bibfnamefont {SeungJung}\ \bibnamefont
  {Huh}}, \bibinfo {author} {\bibfnamefont {Irene~Prieto}\ \bibnamefont
  {Rodríguez}}, \bibinfo {author} {\bibfnamefont {Julian~F.}\ \bibnamefont
  {Wienand}}, \bibinfo {author} {\bibfnamefont {Wolfgang}\ \bibnamefont
  {Ketterle}}, \bibinfo {author} {\bibfnamefont {Markus}\ \bibnamefont {Heyl}},
  \bibinfo {author} {\bibfnamefont {Anatoli}\ \bibnamefont {Polkovnikov}},
  \bibinfo {author} {\bibfnamefont {Immanuel}\ \bibnamefont {Bloch}}, \ and\
  \bibinfo {author} {\bibfnamefont {Monika}\ \bibnamefont {Aidelsburger}},\
  }\href {https://arxiv.org/pdf/2501.16995.pdf} {\enquote {\bibinfo {title}
  {{Probing quantum many-body dynamics using subsystem Loschmidt echos}},}\ }
  (\bibinfo {year} {2025}),\ \Eprint {http://arxiv.org/abs/2501.16995}
  {arXiv:2501.16995} \BibitemShut {NoStop}%
\bibitem [{\citenamefont {Meng}\ \emph {et~al.}(2026)\citenamefont {Meng},
  \citenamefont {Batista},\ and\ \citenamefont {Li}}]{Meng2026}%
  \BibitemOpen
  \bibfield  {author} {\bibinfo {author} {\bibfnamefont {Zi~Yang}\ \bibnamefont
  {Meng}}, \bibinfo {author} {\bibfnamefont {Cristian~D.}\ \bibnamefont
  {Batista}}, \ and\ \bibinfo {author} {\bibfnamefont {Shiliang}\ \bibnamefont
  {Li}},\ }\bibfield  {title} {\enquote {\bibinfo {title} {An integrated
  theoretical and numerical approach to understand modern experiments on
  quantum magnetism},}\ }\href {https://doi.org/10.1038/s41567-026-03304-5}
  {\bibfield  {journal} {\bibinfo  {journal} {Nature Physics}\ }\textbf
  {\bibinfo {volume} {22}},\ \bibinfo {pages} {1189--1197} (\bibinfo {year}
  {2026})}\BibitemShut {NoStop}%
\bibitem [{\citenamefont {Halimeh}\ \emph {et~al.}(2025)\citenamefont
  {Halimeh}, \citenamefont {Aidelsburger}, \citenamefont {Grusdt},
  \citenamefont {Hauke},\ and\ \citenamefont {Yang}}]{Halimeh2025}%
  \BibitemOpen
  \bibfield  {author} {\bibinfo {author} {\bibfnamefont {Jad~C.}\ \bibnamefont
  {Halimeh}}, \bibinfo {author} {\bibfnamefont {Monika}\ \bibnamefont
  {Aidelsburger}}, \bibinfo {author} {\bibfnamefont {Fabian}\ \bibnamefont
  {Grusdt}}, \bibinfo {author} {\bibfnamefont {Philipp}\ \bibnamefont {Hauke}},
  \ and\ \bibinfo {author} {\bibfnamefont {Bing}\ \bibnamefont {Yang}},\
  }\bibfield  {title} {\enquote {\bibinfo {title} {Cold-atom quantum simulators
  of gauge theories},}\ }\href {https://doi.org/10.1038/s41567-024-02721-8}
  {\bibfield  {journal} {\bibinfo  {journal} {Nature Physics}\ }\textbf
  {\bibinfo {volume} {21}},\ \bibinfo {pages} {25–36} (\bibinfo {year}
  {2025})}\BibitemShut {NoStop}%
\end{thebibliography}
\end{document}